\documentclass{aa}  

\usepackage{txfonts}
\usepackage{lipsum}
\usepackage{subcaption}         
                                
\usepackage{lscape}             
                                
\usepackage{placeins}           
                               
\usepackage{natbib}  

\usepackage{graphicx}	
\usepackage[rightcaption, wide]{sidecap}  
\usepackage{amsmath}	
\usepackage{amssymb}	
\usepackage{bm}		    
\usepackage{pdflscape}	
\usepackage{blindtext}
\usepackage{xurl}
\usepackage[hidelinks, breaklinks]{hyperref}
\usepackage{float}
\usepackage{multirow, array}
\usepackage{rotating}
\usepackage[table]{xcolor}
\usepackage{orcidlink}
\usepackage{soul}
\usepackage{afterpage}

\begin{document}
\nolinenumbers

   \title{Diffuse and specular brightness models applied to LEO satellites. Case study: The ONEWEB constellation}
   \titlerunning{Brightness models for the ONEWEB LEO satellites}

   \author{
            Mar\'ia Romero-Colmenares$^{1,7}$ \orcidlink{0000-0001-6515-5036} \thanks{E-mail: maria.romero@postgrados.uda.cl}
            \and Katherine Vieira$^{1}$\orcidlink{0000-0001-5598-8720}\thanks{E-mail: katherine.vieira@uda.cl}
            \and Jeremy Tregloan-Reed$^{2,7}$ \orcidlink{0000-0002-9024-4185}
            \and Yonggi Kim$^{3}$
            \and Joh-Na Yoon$^{3,4}$ 
            \and Ha-eun Kim$^{3,4}$ 
            \and Hyo-ri Jeon$^{3,4}$ 
            \and Chae-rin Kim$^{3,4}$ 
            \and Christian Adam $^{2,7}$ \orcidlink{0000-0002-9021-3928}
            \and Tob\'ias C. Hinse$^{5}$ \orcidlink{0000-0001-8870-3146}
            \and Mario Soto$^{1,7}$ \orcidlink{0000-0001-8444-9742}
            \and Eduardo Unda-Sanzana$^{2,7}$ \orcidlink{0000-0002-7514-8312}
            \and Pen\'elope Longa-Peña$^{2,7}$ \orcidlink{0000-0001-9330-5003}
            \and \'Angel Otarola$^{6}$ \orcidlink{0000-0002-9789-2564}
        }

   \institute{
            Instituto de Astronomia y Ciencias Planetarias, Universidad de Atacama, Copayapu 485, Copiapó 1531772, Atacama, Chile 
            \and Centro de Astronomía (CITEVA), Universidad de Antofagasta, Avenida U. de Antofagasta, 02800 Antofagasta, Chile
            \and Department of Astronomy and Space Science, Chungbuk National University, 28644 Cheongju, Korea
            \and Chungbuk National University Observatory, 28644 Cheongju, Korea
            \and University of Southern Denmark, Department of Physics, Chemistry and Pharmacy, SDU-Galaxy, Campusvej 55, 5230, Odense M, Denmark
            \and European Southern Observatory, Alonso de Córdova 3107, Vitacura, Región Metropolitana, Chile
            \and Chilean Low Earth Orbit satellites (CLEOSat), Chile
            }

   \date{Received XX XX, 20XX}
 
  \abstract
  % context heading (optional)
  % {} leave it empty if necessary  
   {To better understand the observed brightness of low Earth orbit satellites, we must characterize their reflectivity, which in turn depends importantly on their bus designs. The reflectivity of a body can be described by Lambert’s law, in terms of its albedo, cross-sectional area, range (distance), phase angles, and the mixing coefficient between diffuse and specular reflection components.}
  % aims heading (mandatory)
   {We aim to analyze the reflectivity of more than 300 ONEWEB satellites using the diffuse Lambertian sphere, diffuse and specular Lambertian sphere, and the relative reflectance brightness models.} 
  % methods heading (mandatory)
   {Astrometric and photometric measurements, plus two-line elements (TLE) orbital information were used to compute the apparent and range-corrected magnitude, as well as the relevant angles related to the orientation of the Sun, the satellites, and the observer. A differential evolution Monte Carlo algorithm was used to obtain each model's parameters that best fit the data.}
  % results heading (mandatory)
   {All models can fit the mean observed brightness of the satellites but cannot describe the observed phase-angle-dependent brightness modulations. The residuals in all cases have a standard deviation of $\sim$0.6 magnitudes, while the observational photometric errors are on average $\sim$0.2 magnitudes.}
  % conclusions heading (optional), leave it empty if necessary
   {The studied brightness models, which depend on the relative Sun-body-observer position but are independent of the specific orientation of the reflecting body surface(s) with respect to the observer, cannot entirely explain the observed brightness of the ONEWEB constellation satellites. Accounting for the real shape and the changing attitude of the satellite, as well as the effect of Earth's albedo is needed to better explain satellite photometric observations.}


   \maketitle
\nolinenumbers

\section{Introduction}

Companies operating low Earth orbit satellite (LEO) constellations plan to launch tens of thousands of communications satellites over the next decade. Some of these companies are ONEWEB, Starlink, Amazon Kuiper, Guowang, and StarNet \citep{Boley_2022}. The number of satellite launches is continually increasing, with the number of satellites in LEO expected to exceed one hundred thousand by the end of the decade \citep{Venkatesan2020} and over $\sim$\,560\,000 based on requests submitted to the U.S. Federal Communications Commission (FCC) and the International Telecommunication Union (ITU) filings\footnote{See \href{https://planet4589.org/space/con/conlist.html}{\url{ https://planet4589.org/space/con/conlist.html}} accessed 2024-12-17.}. 

Low Earth orbit satellites are generally deployed at altitudes between 350\,km and 2\,000\,km. Depending on the orbital height, the satellites may reflect the sunlight and become visible to observers beyond twilight and into the night. For this reason, satellites are most visible for one to three hours after sunset or before sunrise. Additionally, during summer at high latitudes, LEOsats are visible almost all night \citep{mcdowell2020}.

Due to the number of current and planned LEOsats in orbit, and their ability to reflect sunlight, they negatively impact several areas of observational astronomy, especially the observation of rare transients that require scanning large portions of the sky. Examples of such cases include the search for and tracking of near-Earth objects (NEOs), inner solar system asteroids, as well as deep multi-object spectroscopic surveys such as those performed by the Dark Energy Spectroscopic Instrument (DESI) wide-field spectrograph installed on the Nicholas U. Mayall 4\,m telescope at Kitt Peak National Observatory \citep{DESI} and the Dark Energy Camera (DECam) installed on the 4\,m Victor M. Blanco telescope at the Cerro Tololo Inter-American Observatory in Chile \citep{DECAM}. Satellite brightness might also affect wide-field exoplanet transit surveys such as the Hungarian-made Automated Telescope Network (HATNet), whit its 10.4\,$^\circ$ $\times$ 10.4\,$^\circ$; field of view (FoV) and the Next Generation Transit Survey (NGTS), consisting of 12 telescopes each with an 8\,deg$^2$ FoV, providing a total coverage of 96\,deg$^2$ \citep{NGTS}. Low Earth orbit also reflect in both the near-infrared (NIR) and infrared (IR) regimes; thus, surveys using these wavelengths, including the wide-field NIR camera at the 4\,m United Kingdom Infra-Red Telescope on Mauna Kea \citep{UKIRT-WFC}, are affected.

The most noticeable impact from LEOsats on astronomical observations, in both the optical and the NIR, is bright streaks across the detector. These streaks affect amateur and professional astronomers (see IAU press release, 2019/06/03\footnote{See \href{https://www.iau.org/news/announcements/detail/ann19035/}{\url{ https://www.iau.org/news/announcements/detail/ann19035/}}} and 2020/02/12\footnote{See \href{https://www.iau.org/news/pressreleases/detail/iau2001/}{\url{ https://www.iau.org/news/pressreleases/detail/iau2001/}}}). Different groups in the observational astronomy community have reported the reflective brightness of Starlink (a subsidiary of SpaceX) satellites \citep[e.g.,][]{Tregloan_Reed_2020, Tregloan_Reed_2021, Tyson_2020, Boley_2022, mroz2022, Takashi2023}, while attempting to quantify the risks that LEOsat constellations pose to ground-based astronomy. Recently, an international observational campaign led by the IAU Centre for the Protection of the Dark and Quiet Sky from Satellite Constellation Interference (CPS)\footnote{\href{https://cps.iau.org/}{\url{ https://cps.iau.org/}}} has been successful in measuring the reflective brightness of the AST SpaceMobile Bluewalker 3, a prototype 4G/5G satellite, which at the time of launch was the largest communications satellite in LEO with a 64\,m$^2$ array. The reported results found that BlueWalker 3 shows a stationary V magnitude of $0.4\pm0.1$ \citep{nandakumar2023high}. These results support the importance of the development of national and international legal frameworks, along with recommendations and mitigation strategies, for both LEOsat operators and observatories \citep[Dark \& quiet skies]{Walker2020Impact}.

The impact of individual LEOsats on astronomy depends on how bright they appear. The reflected sunlight from satellites comprises two specular and diffuse components. Specular reflection (also called regular reflection) corresponds to light scattered in a preferential direction (mirror-like reflection), whereas diffuse reflection occurs when light is scattered in several directions. The type of scattering depends on the detailed shape of the satellites. 
 
Both the materials used for the construction of the satellites and the roughness of the surface materials affect the type of scattering. For instance, phased antennas produce a more diffuse scattering effect and less of a specular reflection. This is one reason why the first-generation of Starlink LEOsats, which lacked reflective brightness mitigations, had an on-station visible stationary magnitude of V\,$\sim$\,5, making them visible to the naked eye \citep{Tregloan_Reed_2020, Tregloan_Reed_2021, Walker2020Impact, Tyson_2020, Krantz2023, Boley_2022}. During the satellites' deployment phase, their orientation in space also matters in how they reflect sunlight. \citet{Krantz2023}, and \citet{Boley_2022} studied the effects of the Sun phase angle and range on the apparent magnitude of Starlink satellites using a brightness model: a Lambertian diffuse-specular sphere by \cite{Mccue_1971, Williams1966, Pradhan2019}. Another reason why LEOsats are much brighter than geostationary satellites is because of their low orbits, with altitudes between 350\,km and 1\,200\,km. When considering the inverse square law, a satellite at 500\,km appears 3\,600 times brighter ($\sim 8.9$\,mag brighter) than a satellite at 30\,000\,km. 

To date, several studies have been done to measure the effectiveness of different mitigation designs aimed at reducing the reflective brightness of LEOsats \citep[e.g.,][]{Hainaut_2020, mcdowell2020, Tyson_2020, Tregloan_Reed_2020, Tregloan_Reed_2021}. Studies report that for the first Starlink mitigation design, Darksat, the optical magnitude was reduced from $\sim 5$\,mag to $\sim 6$\,mag while in the NIR it went from $\sim 2$\,mag to $\sim 3$\,mag \citep{Tregloan_Reed_2020, Tregloan_Reed_2021, Tyson_2020}. The impact of reflected sunlight from LEOsats on astronomy will be even greater for telescopes with instruments designed for long exposures on ultra-wide fields, due to the combined effect of the telescope light-collecting area and its angular FoV \citep{Hainaut_2020, mcdowell2020}. A notable example of this is the Charles Simonyi telescope’s camera detectors at the National Science Foundation's Vera C. Rubin Observatory. \citet{Tyson_2020} analyzed the impact of LEOsats on this telescope and camera. They concluded that an optical magnitude of g\,$\geq$\,7 is necessary for satellites located at an orbital height of 550 km to enable corrections of the trails left by LEOsats to a level below the background noise. For orbital heights above 550\,km, the LEOsats must be fainter because their lower angular velocity means they illuminate they illuminate the detector pixels for longer. The first workshop on the issue of satellite constellations and their impact on astronomy \citep{Walker2020Impact} found that the magnitude limit $m_{\mathrm{lim}}$, for which the above-mentioned correction is possible, is related to the orbital height $h$ by
\begin{equation}
m_{\mathrm{lim}} = \begin{cases}
    7 & \text{when $h\leq$ 550\,km}\\
    7 + 2.5 \log_{10}\left(h/550\,km\right) & \text{when $h>$ 550\,km}\\
\end{cases}  \, .
\end{equation}
For example, satellite constellations with orbital heights of 1\,200\, km, such as ONEWEB, must not exceed g\,$=7.9$\,mag.

The characterization of the reflected brightness of an object can be represented by different mathematical models. For example, the \citet{Mccue_1971} model offers an approximation of the reflected brightness of an object with different geometries (spherical, cylindrical, and flat). On the other hand, the \citet{Minnaert1941} model studied how light is reflected on an opaque surface. Both models have been used to characterize observations of Starlink LEOsats by \citet{Pradhan2019, Mallama2020_SL, Tregloan_Reed_2020, Tregloan_Reed_2021, Lawler_2022, Boley_2022, Krantz2023}.

The aim of this work is the characterization of the stationary magnitude of the ONEWEB satellite constellation using two different sphere models by \citet{Mccue_1971} which consider the phase angle,  range, effective area, cross section, albedo and mixing coefficient. In addition, we study the reflective opacity of LEOsats using the approach of \citet{Tregloan_Reed_2020}, based on the model of \citet{Minnaert1941}, which considers the incidence and observer angles and edge darkening parameter. All models are applied to a group of over 300 satellites. 

\begin{figure}[t]
    \includegraphics[clip, trim=0.0cm 0.0cm 0.0cm 0.0cm,width=8cm]{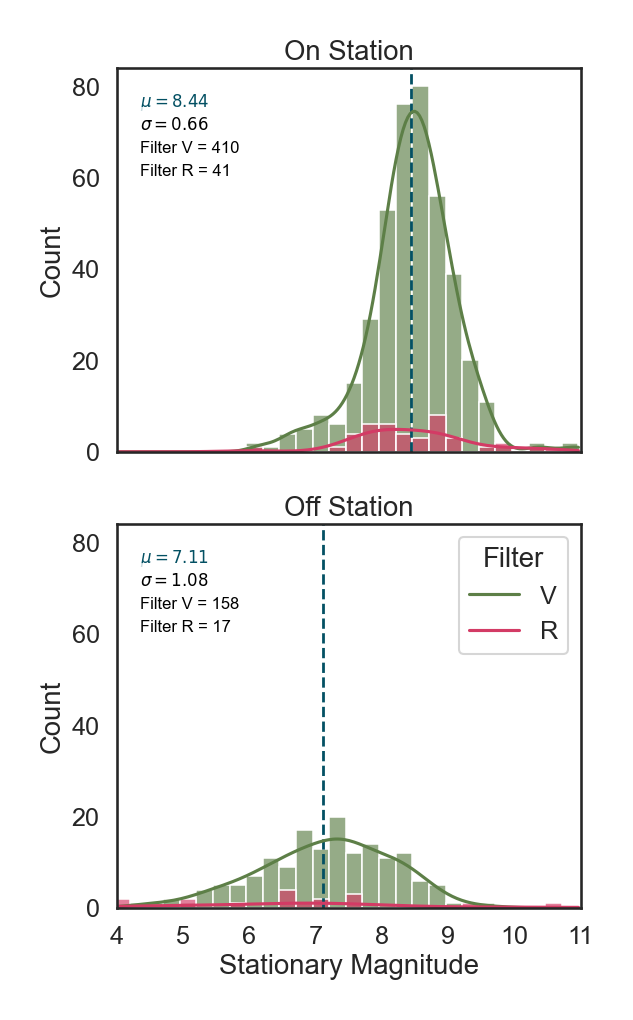}
    \caption{Distribution of stationary magnitudes of all ONEWEB satellites studied. The sample is divided by filter and orbit type. There are 451 on-station satellites, of which 410 were observed with the V filter and 41 with the R filter. There are 175  off-station satellites observed 158 with the V filter and 17 with the R filter. The average $\mu$ and standard deviation $\sigma$ of $m_{s}$ are shown in the legend.}
    \label{fig:distribution_estmag}
\end{figure}

A satellite constellation consists of a group of artificial satellites with identical circular orbits, height and inclination, grouped in a series of orbital planes with nodes located along the equatorial plane, as defined by \cite{walker1984}. The ONEWEB constellation consists of a set of 720 satellites designed and manufactured by Airbus Defence and Space and ONEWEB Ltd\footnote{Accessed 2024-09-10 \href{https://airbusus.com/en/products-services/}{\url{https://www.airbus.com/en/products-services/space/telecommunications-and-navigation-satellites/low-earth-orbit-satellite-constellation}}.}. The satellites have a box-wing design with a central box-shaped bus and two solar panel arrays. The design also includes two extended antenna reflectors \citep{Krantz2023}. The constellation has 18 circular polar-orbital planes at an altitude of 1\,200\,km and inclined is 87$^\circ$, providing global coverage. The main aim of this constellation is to enable high-speed, low-latency internet access worldwide, including remote parts of the world \citep{Portillo_2019}.

The ONEWEB constellation has three different stage orbits (i.e., orbital parked, orbital rise, and on-station), which depend on their activity. Orbital parked is when the satellites have an altitude of less than $\sim$\,630\,km, and they are parked at the start of their service life while waiting to perform an orbital rise or complete service. Orbital rise is when the satellite has an altitude between $\sim$\,630\,km to $\sim$\,1\,100\,km. In this stage satellites are maneuvered to their final orbit while a increasing altitude. On-station is when the satellites have reached their operational height (t$\sim\,$1\,200\,km) and are ready to start operations. A plot of the range versus altitude of the data studied in this paper is shown in Figure \ref{fig_apx:satellite_altitude_distribution} in Appendix \ref{apx:more_figures}. As of 2024, all the ONEWEB constellation satellites occupy their final on-station position.

\section{Observation and data reduction}

To observe LEOsats, it is necessary to know where to point the telescope. This information is encoded in the two-line elements (TLE), which is a standardized format used to describe the orbit elements of a satellite. Two-line elements consist of a set of data that includes orbital parameters, such as the satellite's position, velocity, and other relevant information at a specific epoch or time \citep{croitoru2016}. Two-line elements were developed by the North American Aerospace Defense Command (NORAD). NORAD is an organization established by the United States and Canada to monitor North American airspace for objects in orbit between $\sim100$\,km, and 40\,000\,km altitude. All TLEs are displayed on a website where they are available to the public. They can be used for multiple technical studies when the position of a satellite is needed at some instant in time.

\renewcommand{\arraystretch}{1.5}
\begin{table*}[h]
    \begin{center}
        \caption{Sample of the results provided by the \texttt{Satellite-Tracking} code.}
        \begin{tabular}{c>{\columncolor[HTML]{EFEFEF}}cc>{\columncolor[HTML]{EFEFEF}}cc}
        \hline
        \multicolumn{1}{c}{ID Satellite} & \multicolumn{1}{>{\columncolor[HTML]{EFEFEF}}c}{Date} & \multicolumn{1}{c}{Time}& \multicolumn{1}{>{\columncolor[HTML]{EFEFEF}}c}{RA}              & \multicolumn{1}{c}{DEC}           \\ 
        \multicolumn{1}{c}{ } & \multicolumn{1}{>{\columncolor[HTML]{EFEFEF}}c}{ (UT)} & \multicolumn{1}{c}{UT}& \multicolumn{1}{>{\columncolor[HTML]{EFEFEF}}c}{hh:mm:ss}              & \multicolumn{1}{c}{dd:mm:ss}           \\ \hline 
        ONEWEB-0010 & 2022 05 01 & 12 33 30 & 10 51 42.389 & 48 39 02.24 \\ 
        ONEWEB-0011 & 2022 05 18 & 12 30 30 & 14 50 01.642 & 25 01 51.84 \\ 
        ONEWEB-0461 & 2022 04 21 & 12 12 30 & 06 19 17.347 & 50 17 54.65 \\ 
        ONEWEB-0473 & 2022 04 21 & 11 53 30 & 11 23 18.129 & 23 47 43.83 \\ 
        ONEWEB-0474 & 2022 04 21 & 12 07 30 & 10 15 04.469 & -13 32 01.02 \\ \hline
        \label{tab:tle_transformation}
        \end{tabular}
    \end{center}
\end{table*}

The task of transforming the data provided by the TLEs into celestial coordinates was carried out using the \texttt{Satellite-Tracking}\footnote{Available from the CLEOsat Group \href{https://github.com/LEO-Satellites/satellite-tracking}{\url{https://github.com/LEO-Satellites/satellite-tracking}} code, developed by authors AO, JTR and Edgar Ortiz.} code. To use the code, users must select the observatory and the observation time (evening or morning twilight). Then, the \texttt{Satellite-Tracking} code downloads the latest available TLE from the Celestrak\footnote{\href{https://celestrak.org/NORAD/elements/supplemental/}{\url{https://celestrak.org/NORAD/elements/supplemental/}}} website, outputs the results in an easy-to-read text file as a function of time, allowing observers to scroll down the list, selecting the targets to observe, as seen in Table \ref{tab:tle_transformation}.

In general, when observing LEOsats, the wait-and-capture method is used, which consists of tracking background stars at sidereal speed. This type of observation produces an image where the stars are dots and the LEOsats are represented by a line crossing the image. The angular velocity of LEOsats depends on their orbital height and sky position (their angular velocity decreases at higher elevations), and it can range from a few arcmin\,s$^{-1}$ to about $2^\circ$\,s$^{-1}$. This effectively means that LEOsats can transverse the FoV in less than a second (depending on the angular velocity and FoV size), which requires precise timing of the opening and closing of the camera shutter. During exposures of a few seconds, the light collected from the LEOsat must be scaled to the exposure time, as sky and stellar light are effectively collected throughout, whereas the satellite's full trail may partially fall outside the FoV. This is expressed in equation (\ref{eq:stationary_mag}). 

Data reduction and astrometric calibration were performed using the \texttt{LEOSatpy}\footnote{\href{https://github.com/CLEOsat-group/leosatpy}{\url{https://github.com/CLEOsat-group/leosatpy}}, developed by Christian Adam.} code. The code was written in Python and uses Astropy for flat-field, bias, and dark subtraction. The astrometric calibration begins with the alignment of images with the same background, both with and without satellites, followed by the identification of references stars from the Gaia Early Data Release 3 catalog \citep{Gaia_colla}. 

The next step is to compute the stationary magnitude $m_s$ and position of the satellite at the midpoint of the line. To this end, the magnitude of each LEOsat is extrapolated using the equation by \citet{Boley_2022}: 
\begin{equation}
    m_{s}=m_{0} + m_{\mathrm{ins}} - 2.5\log_{10}\left( \frac{\phi}{L} t_{\mathrm{exp}}\right) ,
     \label{eq:stationary_mag}
 \end{equation}
 where $m_0$ is the zero-point magnitude of the stars, $m_{\mathrm{ins}}$ is the instrumental aperture magnitude, $\phi$ is the angular velocity of the satellite in the sky, $L$ is the length of the line left by the LEOsat, and $t_{\mathrm{exp}}$ is the exposure time. The stationary magnitudes distribution of our sample of LEOsats is shown in Figure \ref{fig:distribution_estmag}.

After that, the range-corrected magnitude $H^{1200}$ is set as the brightness that the LEOsat would have at a range of 1\;200 km, the nominal altitude of the ONEWEB constellation, as follows:
\begin{equation}
H^{1200} = m_s - 5\log_{10}\left(\frac{r}{1\,200\mbox{ km}}\right) ,
     \label{eq:h1200_mag}
 \end{equation}
where $r$ is the range of the satellite in km. The observations used in this work were made using a 60 cm-aperture telescope operated by the Chungbuk National University Observatory in Jincheon (CbNUOJ) in South Korea \citep{Kim2014, Wang2019}. The telescope used an SBIG STX-16083 charge-coupled device (CCD) camera of 4\,096\,$\times$\,4\,096 pixels, allowing a FoV of 72'\,$\times$\,72' and a pixel scale of 1.05\,arcsec/pixel. In addition, Johnson/Cousins UBVRI filters can be used through an electronic filter wheel system. This study uses images taken with V and R filters with a binning of 2\,$\times$\,2 pixels to compensate for the altitude of the ONEWEB satellites by collecting more light. 

The telescope control system uses an internet program that connects to a Yb (Ytterbium) optical lattice clock at KRISS (Korea Research Institute of Standards and Science, \citealt{KRISS}) to synchronize the time each minute. The precision of the time synchronization is $\sim\pm\,0.01$\,s. In addition, the precision of CCD shutter operation is measured to $\pm\,0.005$\,s.

The data presented in this work consist of 626 observations of 322 individual ONEWEB satellites, observed over 41 nights (from October 2021 to September 2022) during evening twilight. A total of 175 observations were conducted for off-station satellites (579 to 1\,100\,km) whose average altitude is 684.94$\pm$108.79 km and average range is 914.37$\pm$182.01 km, and 451 observations of on-station satellites (1\,100 to 1\,229\,km) whose average altitude is 1211.40$\pm$9.05 km and average range is 1542.20$\pm$201.87 km, as shown in Figure \ref{fig_apx:satellite_altitude_distribution}. 

The data were studied as a whole and also separated into four samples according to the filter used (V or R) and their on- or off-station position to study how the brightness models behave for satellites at different altitudes. More details about the samples per filter can be seen in Table \ref{tab:stad_satellite_altitud}. Appendix \ref{apx:glossary} contains a glossary of all the variables and notations used in this work.

\section{Modeling of the satellite brightness}\label{sec:modeling}

Satellites shine because they reflect sunlight, i.e., sunlight hits the satellites and their parts (the main body, antennas, and solar panels), which reflect a fraction of this sunlight toward our direction. This sunlight reflected by the satellite is what we can observe from Earth and measure through astronomical observations by the stationary magnitude $m_s$. The satellite does not reflect all the light it receives, and the amount of light we observe depends on several factors:
\begin{itemize}
    \item[(i)] the physical characteristics of the satellite: its shape and those of its parts (main body, antennas, and solar panels), the type of scattering (diffuse or specular), the materials they are made of, and the roughness of the surfaces;
    \item[(ii)] the geometry between the Sun, the satellite, and the observer;
    \item[(iii)] the orientation of the satellite itself, its solar panels, and antennas with respect to the observer;
    \item[(iv)] the distance between the satellite and the observer. 
\end{itemize}
In practice, no function can describe all these factors, but it is necessary to find models that can estimate most of them. The approach is to assume several hypotheses that allow to simulate the general trends and brightness distributions of satellites. 

In this work, we consider two types of models: the Lambertian spherical models which allow the observed magnitude to be reproduced as a function of the Sun-satellite-observer geometry and type of scattering (diffuse or specular), and the relative reflectance (RR) model which compares the magnitudes observed from two different orientations. These models are generally referred to in the literature as bidirectional reflectance distribution functions or BRDFs \cite[p.~26]{Shepard2017}. 

\subsection{Lambertian spherical models}
In this model, we consider the satellites as having a uniform spherical shape. In addition, we consider two cases of dispersion: one in which the uniform sphere is diffuse, and another in which the uniform sphere is both diffuse and specular. The specular reflection corresponds to light scattered in one direction as in a mirror, whereas diffuse reflection is when light is scattered in several directions.

The stationary magnitude $m_s$ of LEOsats can be reproduced using the Lambertian sphere model described by \citet{Williams1966} as follows:
\begin{equation}
    m_s = m_\odot - 2.5 \cdot \log_{10}[\zeta \cdot F_0(\Phi)] + 5 \cdot \log_{10}(r) \; ,
    \label{eq:lambertian_model}
\end{equation}
where $m_{\odot}=-26.77$ is the Sun's apparent magnitude in the Johnson V filter \citep{Willmer_2018}, 
$\zeta$ is the so-called effective area of the sphere reflecting light from the Sun, 
$F_{0}(\Phi)$ is the so-called phase function which describes the type of scattering mentioned above 
as a function of the Sun phase angle $\Phi$, and $r$ is the range of the satellite (distance between it and the observer). $\zeta=\rho\cdot A$, where $\rho$ is the albedo of the reflecting sphere and $A=\pi R^2$ is its cross-area. The Sun phase angle $\Phi$ is defined as the angle between the Sun, the satellite, and the observer (see Fig. \ref{fig:angles_draw}). $\Phi$ is measured on the plane formed by two vectors: the one going from the satellite to the Sun and the one from the satellite to the observer. Of the several factors affecting the observed brightness and itemized at the beginning of Section \ref{sec:modeling}, in equation (\ref{eq:lambertian_model}), $\Phi$ accounts for the effects of item (ii), $\zeta$ likewise for item (i), while the $\log_{10}(r)$ term measures the effect of item (iv). Appendix \ref{apx:lambertian} briefly describes the derivation of this equation.

\subsubsection{Lambertian diffuse spherical  model}\label{sec:difuse_model}
The diffuse spherical (DS) model assumes that a LEOsat reflects light like a uniform diffuse sphere, in which case the phase function is given by 
\begin{equation} 
    F_{0}(\Phi)  =  \frac{2}{3\cdot\pi^{p+1}}  \cdot \left[(\pi - \Phi)\cdot \cos(\Phi) + \sin(\Phi)\right]^{p}\; .
    \label{eq:f0_ds}
\end{equation}
\citet{Boley_2022} has used this model.

\subsubsection{Lambertian diffuse and specular spherical  model}\label{sec:diffuse_specular_model}
The diffuse and specular spherical (DSS) model assumes that a LEOsat reflects light like a uniform diffuse and specular sphere, in which case the phase function is given by
\begin{equation} 
    F_{0}(\Phi) = \beta F_{1}(\Phi) + (1-\beta)F_{2}(\Phi)\; ,    \label{eq:f0_dss}
\end{equation}
with $0\leq \beta\leq 1$ being the mixing coefficient of diffuse-specular reflection. 
The function $F_{1}(\Phi)$ represents the diffuse part of the phase function, while $F_{2}(\Phi)$ represents the specular portion of it, given respectively by
\begin{eqnarray}
    F_{1}(\Phi) &=& \frac{2}{3\cdot\pi^{2}} \big[(\pi-\Phi)\cdot \cos(\Phi) + \sin(\Phi)\big] \;, \label{eq:function_F1} \\
    F_{2}(\Phi) &=& \frac{1}{4\pi}. \label{eq:function_F2}
\end{eqnarray}
When $\beta=0$, $F_{0}(\Phi)$ is only specular, and when $\beta=1$, $F_{0}(\Phi)$ is only diffuse and equivalent to equation (\ref{eq:f0_ds}) for $p=1$.
\citet{Williams1966, Mccue_1971, Lawler_2022, Krantz2023} have used this model. 

In the Lambertian spherical models, both $A$ and $\rho$ shift the model up or down (larger area or albedo, brighter magnitudes, and reciprocally). For the DS model, $p$ changes the slope of the function, which is always concave down as a function of $\Phi$. On the other hand, in the DSS model, $\beta$ changes both the slope and concavity of the model. The larger the specular portion, the larger the change in the observed magnitude with $\Phi$, in the sense that a mirror looks brighter than an opaque surface. Also, at low values of $\Phi$ the sphere becomes brighter (sphere is in ``full phase'') and at large values of $\Phi$, it becomes fainter (sphere is in ``new phase'').
In Figures \ref{fig_apx:ds_model_by_p_A_rho} and \ref{fig_apx:dss_model_by_beta_A_rho} in Appendix \ref{apx:more_figures}, we show the behavior of $m_s$ for both models, considering different values of $\rho$, $A$, and $p$ (DS); and $\rho$, $A$ and $\beta$ (DSS), for $r=1\,200$ km. Since in both cases, $\rho$ and $A$ have the same functional form, they cannot be fitted separately; therefore, we fit $\zeta=\rho \cdot A$.

\subsubsection{Effective albedo for the DS and DSS models}

The DS and DSS models are a good approximation for the apparent brightness of satellites, but they do not take into account the actual geometry of the satellite. The apparent brightness of satellites is not only the result of the combination of the reflection of their diffuse and specular parts, but also of small changes in the geometry of the illuminated surface due to the orientation of the satellite with respect to the observer \citet{Krantz2023}. The orientation of the satellite only allows it to reflect light with some of its parts \citep{Williams1966, Mccue_1971}. 

These changes in the reflected light due to the orientation changes can be represented by a changing albedo for a fixed cross section area $A=1$ square meter. Such albedo for $A=1$ is called effective albedo and does not represent a real albedo. In fact, it can be larger than 1, but it represents the relative change between the different geometries of the incident light on the satellite and the reflection produced by its surfaces, i.e., item (iii) listed in Section \ref{sec:modeling}. 

For the DS model, with $p=1$, using equations (\ref{eq:lambertian_model}) and (\ref{eq:f0_ds}), the effective albedo is given by 
\begin{equation}
\rho_{\text{eff,ds}}= \frac{
     3\,\pi^{2} \cdot r^{2} \cdot 10^\frac{m_{\odot}-m_{s}
     }{2.5}}{
     2\cdot[(\pi-\Phi) \cdot \cos\Phi+\sin\Phi]}\; ,
    \label{eq:effective_albedo_ds}
\end{equation}
which has been studied by \citet{Krantz2023}. In this work, we also examine the effective albedo
of the DSS model, with $\beta=0.5$, which from equations (\ref{eq:lambertian_model}, \ref{eq:f0_dss}, \ref{eq:function_F1}) and (\ref{eq:function_F2}) is given by
\begin{eqnarray}
	\rho_{\text{eff,dss}} 
	&=& \frac{ 8 \cdot 3 \cdot \pi^2 \cdot r^{2} \cdot 10^\frac{m_{\odot}-m_{s}}{2.5}}
    {4\cdot 2 \cdot [(\pi-\Phi) \cdot \cos\Phi+\sin\Phi] +3\pi}
    \; . \label{eq:effective_albedo_dss}
\end{eqnarray}

\subsection{Relative reflectance model}

\begin{figure}[t!]
    \includegraphics[clip, trim=0.5cm 0.5cm 1cm 0cm,width=3.1in]{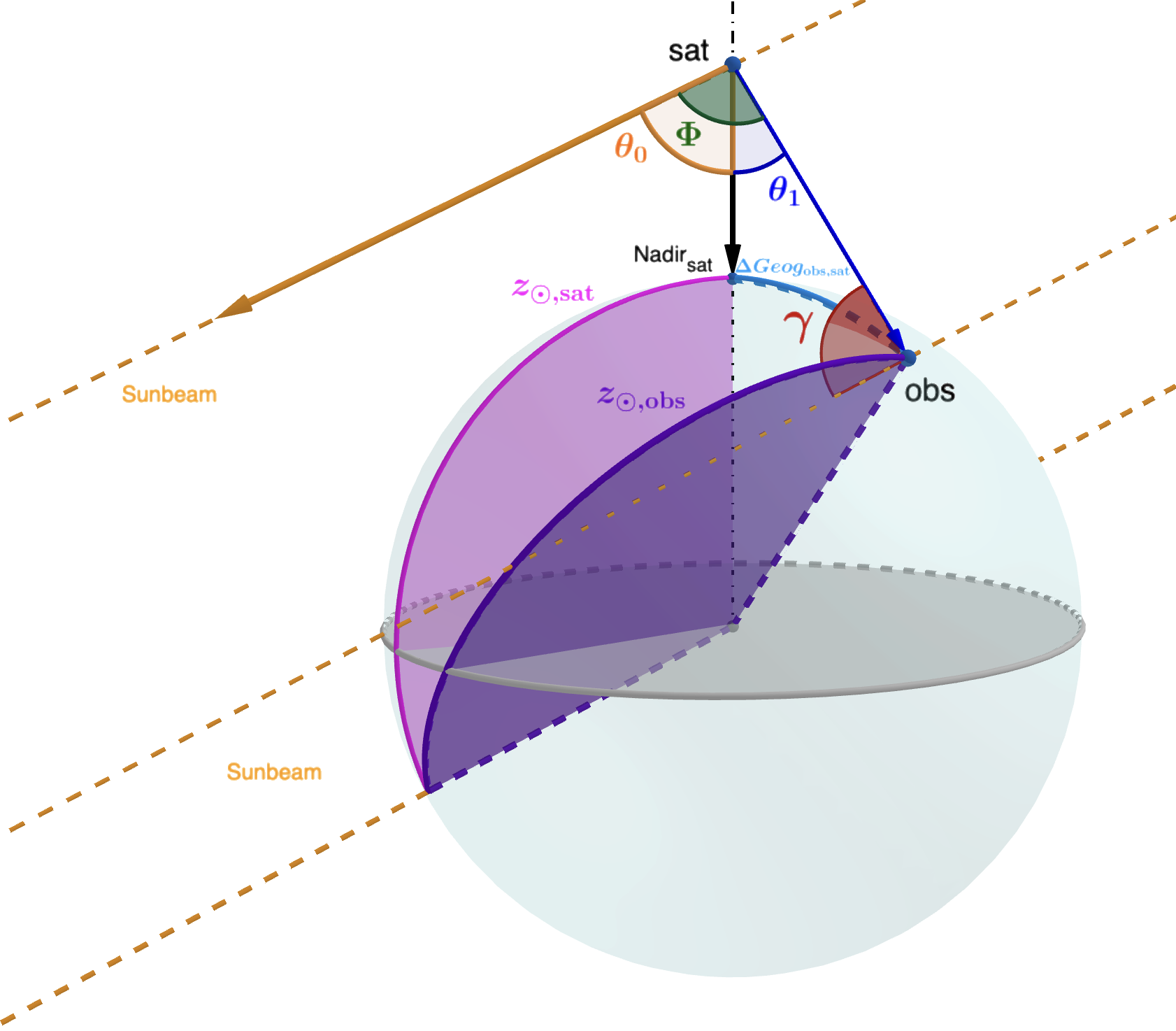}
    \caption{Graphical representation of the position of the satellite and definitions of the Sun incidence angle $\theta_{0}$ (in orange), the observer angle $\theta_{1}$ (in dark blue), the Sun phase angle $\Phi$ (in green), and the solar elongation angle $\gamma$ (in brown). Notice that $\theta_{0}$, $\theta_{1}$ and $\Phi$ are not contained in the same plane but form a triangular pyramid. The distance is not to scale.}
    \label{fig:angles_draw}
\end{figure}

The RR model allows us to compare the magnitudes of a satellite observed at the same range from two different orientations, following \citet{Tregloan_Reed_2021}. Each orientation is parameterized by the angles $(\theta_0,\theta_1)$ and $(\hat{\theta}_0,\hat{\theta}_1)$, where $\theta_0$, and $\hat{\theta}_0$ are the Sun incidence angles and $\theta_1$ and $\hat{\theta}_1$ are the observer angles in each orientation (see Fig. \ref{fig:angles_draw}).
By definition, the difference in stationary magnitudes between two different orientations can be obtained from the ratio of the reflected fluxes by
\[
m_s(\theta_0,\theta_1)-m_s(\hat{\theta}_0,\hat{\theta}_1)
= -2.5  \log_{10}\left( \frac{F_{\theta_0,\theta_1}(r)}{F_{\hat{\theta}_0,\hat{\theta}_1}(r)} \right) \; .
\]
In the RR model, such ratio is
\[
\frac{F_{\theta_0,\theta_1}(r)}{F_{\hat{\theta}_0,\hat{\theta}_1}(r)}=\mathcal{R} = \left( \frac{\cos\;\theta_{0} \times \cos\;\theta_{1}}{\cos\;\hat{\theta}_0 \times \cos\;\hat{\theta}_1} \right)^{k-1} \; ,
\]
where $0\leq k \leq 1$ is known as the edge darkening parameter, which describes the angular distribution of the brightness of the reflected light. When $k=1$, reflectance in all directions is the same, that is the model describes the reflectance of the surface as uniform, i.e., the surface is diffuse \citep{Tregloan_Reed_2021}. 

The ratio $\mathcal{R}$ is the same regardless of the satellite's properties and its range. In other words, the specific values of the stationary magnitudes being compared will change with $r$, the RR model only describes their difference. It is reasonable to assume that a satellite has the same stationary magnitude at a fixed reference orientation and range; therefore, we can convert any observed $m_s$ to that fixed value. Consequently, with the RR model, we reproduce the observed range-corrected magnitude $H^{1200}$, using the following equation:
\begin{equation}
    H^{1200}_{(\theta_0,\theta_1)} = H^{1200}_{(70,0)}
                    - 2.5\log_{10} \biggl[\biggl(
        \frac{\cos\;\theta_{0} \times \cos\;\theta_{1}}  
        {\cos\;70^\circ \times \cos\;0^\circ}
        \biggl)^{k-1}\biggl] \;,
   \label{eq:RR}
\end{equation}
where we chose the reference orientation $(70^\circ,0^\circ)$ and $H^{1200}_{(70,0)}$ is an additional parameter of the model to estimate. The chosen $\hat{\theta}_0=70^\circ$ corresponds to the Sun at $30^{\circ}$ below the horizon and $\hat{\theta}_1=0^\circ$ for the satellite at the observer's zenith. This brightness model attempts to describe the relative effects of item (iii) in Section \ref{sec:modeling}.

Figure \ref{fig:angles_draw} shows a graphical representation of the angles used in the brightness models: Sun phase angle $\Phi$, Sun incidence angle $\theta_{0}$, and observer angle $\theta_{1}$. These angles are defined as follows: $\Phi$ is formed by vectors $(\overrightarrow{\text{SatSun}})$ and $(\overrightarrow{\text{SatObs}})$, $\theta_{0}$ is formed by $(\overrightarrow{\text{SatSun}})$ and $(\overrightarrow{\text{SatNadir}_{\text{Sat}}})$, and $\theta_{1}$ is formed by $(\overrightarrow{\text{SatNadir}_\text{Sat}} )$ and $( \overrightarrow{\text{SatObs}})$.
The Sun incidence angle is obtained from $\theta_{0} = 180^{\circ} - z_{\odot,\text{sat}}$, where $z_{\odot,\text{sat}}$ is the Sun's zenith distance at the Satellite, estimated from the expression:
\begin{eqnarray*}
    \cos(z_{\odot,\text{sat}})\,&=&\,\cos(\Delta \text{Geog}(\text{obs},\text{sat}))\cdot \cos(z_{\odot,\text{obs}}) +  \\
    && \sin(\Delta \text{Geog}(\text{obs},\text{sat})) \cdot \sin(z_{\odot,\text{obs}}) \cdot \\
    && \cos(\Delta Az(\odot,\text{sat})) \; ,
\end{eqnarray*}
where $\Delta \text{Geog}(\text{obs},\text{sat})$ is the angular distance between observer's longitude and latitude and the satellite's longitude and latitude, $\Delta Az(\odot,\text{sat})$ is the difference between the azimuth of the satellite and the azimuth of the Sun, and $z_{\odot,\text{obs}}$ is the Sun's zenith distance for the observer. 

On the other hand, the observer angle $\theta_{1}$ is obtained from the triangle $\bigtriangleup O_{\oplus}O_{\text{sat}}O_{\text{obs}}$:
\begin{eqnarray*}
R_\oplus^2 &=& r^2 + (R_\oplus + H_\text{sat})^2 - 2\cdot r\cdot  (R_\oplus + H_\text{sat})\cdot\cos \theta_1 \Longrightarrow \\
\theta_1 &=& \arccos\left(\frac{r^2 + 2 \cdot R_\oplus \cdot H_{\text{sat}} + H_{\text{sat}}^2}{2 \cdot r \cdot (R_\oplus+H_{\text{sat}})}\right)
\end{eqnarray*}
where $R_\oplus$ is the Earth's radius, $H_{\text{sat}}$ is the perpendicular height (above Earth) of the satellite, and $r$ is the range of the LEOsat. The angles $\theta_0$ and $\theta_1$ are generally not in the same plane, due to a third direction or vector (from the satellite to its nadir) considered in their definitions. 

The solar elongation angle $\gamma$ is obtained from the law of cosines on the equatorial coordinates of the Sun $(\alpha_{\odot}, \delta_{\odot})$ and satellite $(\alpha_{\text{sat}}, \delta_{\text{sat}})$:
\begin{equation}
    \cos(\gamma) = \sin(\delta_{\odot}) \sin(\delta_{\text{sat}}) + \cos(\delta_{\odot})\cos(\delta_{\text{sat}})\cos(\alpha_{\odot} - \alpha_{\text{sat}})
    \label{eq:gamma}
\end{equation}
Finally, to compute the Sun phase angle $\Phi$, we must notice that since the Sun can be taken at being at an infinite distance; therefore, the directions satellite-Sun and observer-Sun are parallel and $\Phi =180^{\circ} - \gamma$.

\section{TLE-predicted versus actual position of satellites}

Two-line elements have some limitations in terms of the accuracy of the data they provide. The Simplified General Perturbation 4 (SGP4) model used to define the orbital parameters includes the dynamic effects of atmospheric drag and solar radiation pressure, providing a fast propagation time but low accuracy (e.g., \citet{Acciarini2025}, their Section 4.1). Furthermore, TLEs do not consider maneuvers made by operating companies to adjust the orbit. In addition, the TLE format does not include covariance matrices that allow error estimation from the orbital parameters \citep{Vallado2012}. 

One way to estimate the accuracy of TLE positions is to use satellite observations to estimate these errors. Astronomical observations are independent of the data and the SGP4 model that generates the TLEs. The results of the \texttt{Satellite-Tracking} code tell us the time and equatorial coordinates at which we must point the telescope to see a satellite pass through the center of the image. However, in most cases, the satellite does not pass through the center of the image, and there is an angular difference $\sigma_{\text{tle}}$ between the TLE-predicted and the actual observed coordinate. 

In our sample, these differences are within 2 arc minutes, which is around the typical value observed by other authors with other LEOsats constellations, --for example \citet[Blue Walker, $7.2\pm 3.1$ arcmin]{nandakumar2023high}, \citet[ONEWEB constellation, between 1.8 and 2.4 arcmin]{Krantz2023}, and \citet[Starlink constellation, on average 0.576 deg]{Halferty2022}.  

\begin{figure}[t]
    \includegraphics[ trim=0.5cm 0.5cm 1cm 0cm,width=3in]{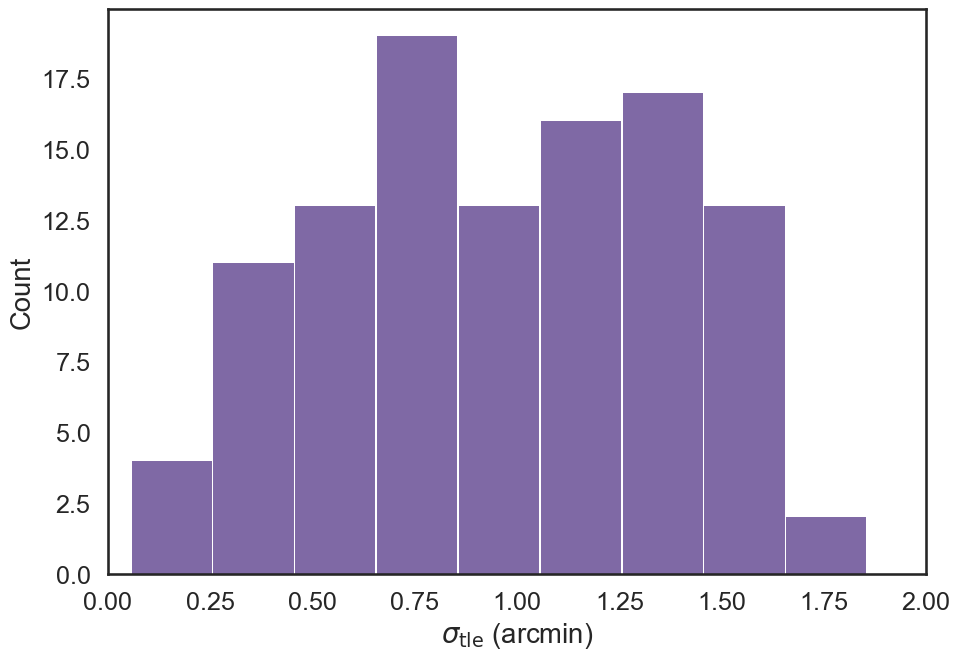}
    \caption{Histogram of $\sigma_\text{tle}$ for all the ONEWEB satellites studied, showing the difference between the TLE-predicted and the observed equatorial coordinated of satellites.}
    \label{fig:tle_errors}
\end{figure}

\renewcommand{\arraystretch}{1.5} 
\begin{table*}[t]
    \begin{center}
    \caption{DEMC specifications.}
    \begin{tabular}{c>{\columncolor[HTML]{EFEFEF}}c>{\columncolor[HTML]{EFEFEF}}c>{\columncolor[HTML]{EFEFEF}}cc>{\columncolor[HTML]{EFEFEF}}c}
                           & \multicolumn{3}{>{\columncolor[HTML]{EFEFEF}}c}{Stochastic random variables}                & Deterministic random variable & Likelihood of observations                                        \\ \hline
\multirow{2}{*}{DS Model}  & $p$       & $\zeta$             & $\sigma_{\text{demc,ds}}$  & $\mu_{\text{demc,ds}}$       & Y\_obs                                                            \\
                           & $U(0, 1)$ & $U(0, 1)$           & $HN(0, 1^2)$                 & DS function                   & $m_s-\mu_{\text{demc, ds}} \sim N(0, \sigma^2_{\text{demc,ds}})$   \\ \hline
\multirow{2}{*}{DSS Model} & $\beta$   & $\zeta$             & $\sigma_{\text{demc,dss}}$ & $\mu_{\text{demc,dss}}$      & Y\_obs                                                            \\
                           & $U(0, 1)$ & $U(0, 1)$           & $HN(0, 1^2)$                 & DSS function                  & $m_s-\mu_{\text{demc, dss}}\sim N(0, \sigma^2_{\text{demc,dss}})$ \\ \hline
\multirow{2}{*}{RR Model}  & $k$       & $H^{1200}_{(70,0)}$ & $\sigma_{\text{demc,rr}}$  & $\mu_{\text{demc,rr}}$       & Y\_obs                                                            \\
                           & $U(0, 1)$ & $U(5, 12)$          & $HN(0, 1^2)$                 & RR function                   & $H^{1200}-\mu_{\text{demc,rr}}\sim N(0,\sigma^2_{\text{demc,rr}})$ \\ \hline
    \end{tabular}
    \label{tab:demc_parameters}
   \tablefoot{For each model (first column), the prior distributions assumed for the parameters $p$ and $\zeta$, and magnitude noise ($\sigma_{\text{demc}}$) (second column), the function being fitted (third column), and the likelihood of the observations (last column). $U$, $N$, and $HN$ mean uniform, normal, and half-normal distributions, respectively.}
    \end{center}
\end{table*}

\renewcommand{\arraystretch}{2} 
\begin{table*}[t]
    \begin{center}
    \caption{DEMC-fitted parameters and errors of the DS, DSS and RR brightness models.}
    \begin{tabular}{ccc>{\columncolor[HTML]{EFEFEF}}cc>{\columncolor[HTML]{EFEFEF}}c}
                            &                            && All      & On Station        & Off Station       \\ \hline 
                            & $p$                        && 0.351\;$\pm$\;0.042 & 0.422\;$\pm$\;0.045 & 0.266\;$\pm$\;0.090 \\ 
                            & $\zeta$                    &$(m^2)$& 0.125\;$\pm$\;0.005 & 0.124\;$\pm$\;0.005 & 0.136\;$\pm$\;0.014 \\
\multirow{-3}{*}{DS Model}  & $\sigma_{\text{demc,ds}}$     &(mag)& 0.734\;$\pm$\;0.021 & 0.635\;$\pm$\;0.022 & 0.890\;$\pm$\;0.048 \\ \hline
                            & $\beta$                    && 0.222\;$\pm$\;0.006 & 0.205\;$\pm$\;0.006 & 0.266\;$\pm$\;0.017 \\
                            & $\zeta$                    &$(m^2)$& 0.383\;$\pm$\;0.041 & 0.448\;$\pm$\;0.044 & 0.297\;$\pm$\;0.090 \\ 
\multirow{-3}{*}{DSS Model} & $\sigma_{\text{demc,dss}}$     &(mag)& 0.751\;$\pm$\;0.021 & 0.685\;$\pm$\;0.023 & 0.879\;$\pm$\;0.047 \\ \hline
                            & $k$                        && 0.542\;$\pm$\;0.073 & 0.526\;$\pm$\;0.078 & 0.368\;$\pm$\;0.168 \\
                            & $H^{1200}_{(70,0)}$        &(mag)& 7.694\;$\pm$\;0.041 & 7.759\;$\pm$\;0.042 & 7.422\;$\pm$\;0.110 \\
\multirow{-3}{*}{RR Model}  & $\sigma_{\text{demc,rr}}$     &(mag)& 0.752\;$\pm$\;0.021 & 0.685\;$\pm$\;0.023 & 0.879\;$\pm$\;0.047 \\ \hline
    \end{tabular}
    \label{tab:all_models}
    \end{center}
    
\end{table*}

\section{Data analysis: Differential evolution Monte Carlo} \label{sec:data_analysis}

We start by looking at the stationary magnitude distribution of the ONEWEB satellites, as obtained from equation (\ref{eq:stationary_mag}) and displayed in Figure \ref{fig:distribution_estmag}. Our data fall within the range presented by \citet{Krantz2023}, with $3.9\leq m_s\leq 11.0$ mag. The average magnitude for the whole sample is 7.8 mag, in which 74$\%$ are on-station satellites and 26$\%$ off-station satellites. The mean stationary magnitude error is 0.19 with a standard deviation of 0.06, and it ranges between 0.1 and 0.3 magnitudes for 98\% of the data.

\begin{SCfigure*}[1.5][h!]
        \includegraphics[width=0.7\textwidth]{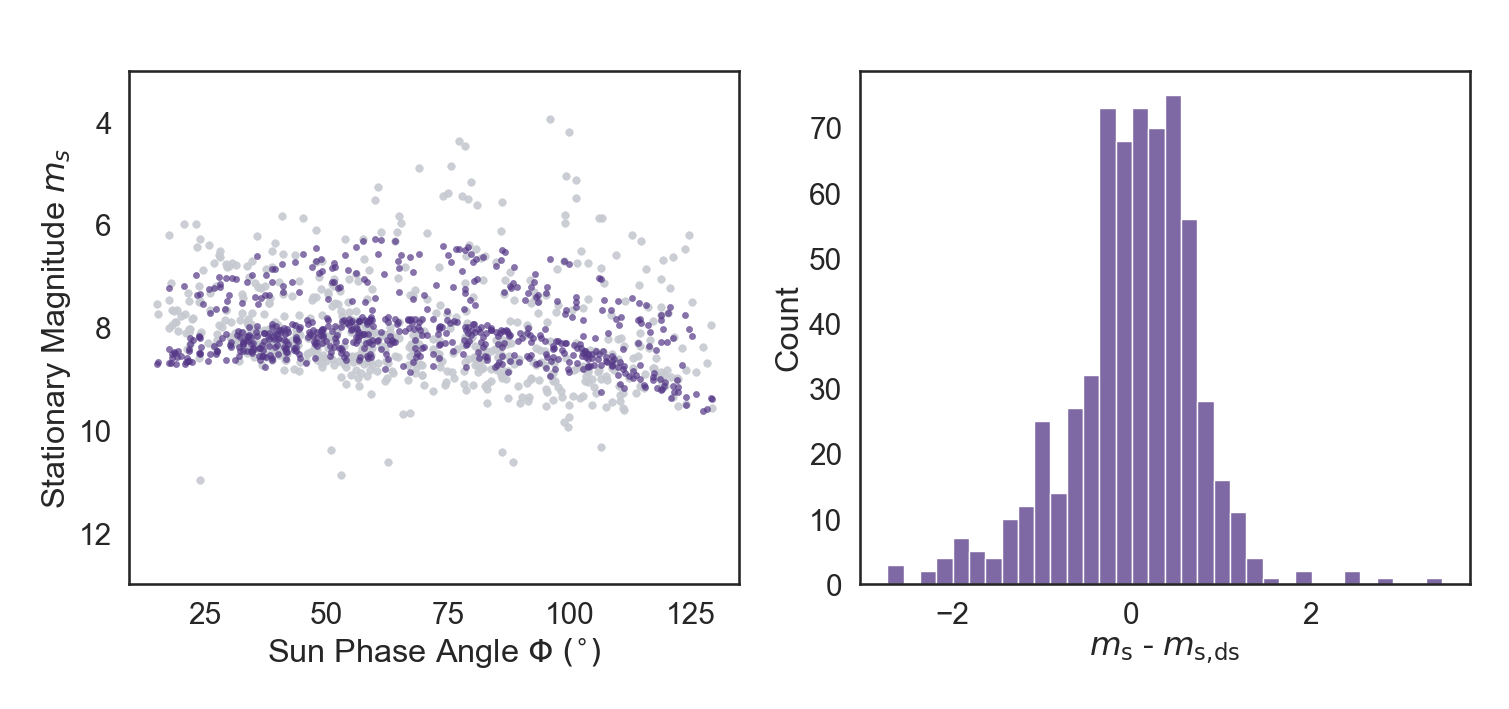}
        \caption{Results of the DS model applied to all the ONEWEB LEOsats of this study. Left: Observed $m_s$ (gray points) and DS model estimates $m_{s,\text{ds}}$ (purple points) vs. $\Phi$. Right: Histogram of residuals $m_s-m_{s,\text{ds}}$.}
        \label{fig:ds_estmag_all}
\end{SCfigure*}

\begin{SCfigure*}[1.5][h!]
        \includegraphics[width=0.7\textwidth]{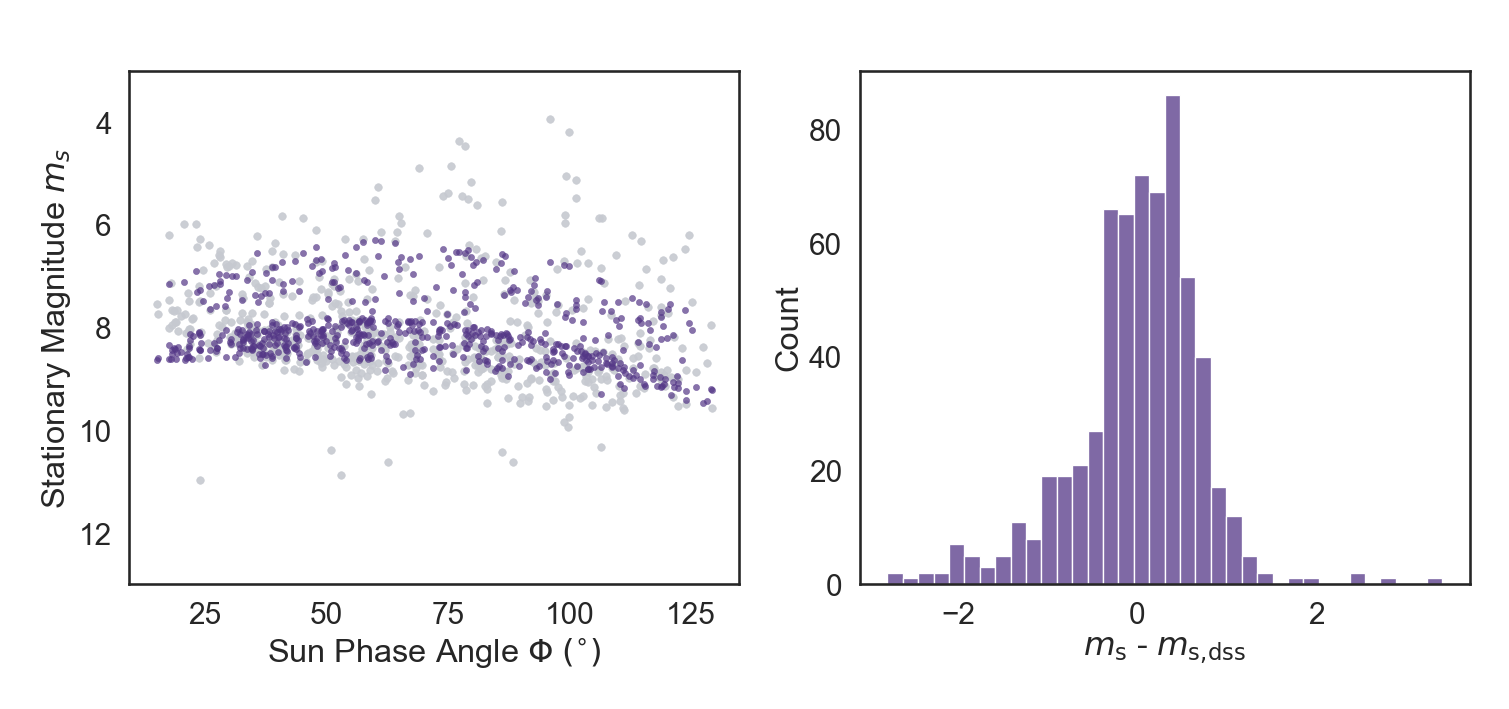}
        \caption{Results of the DSS model applied to all the ONEWEB LEOsats of this study. Left: Observed $m_s$ (gray points) and DSS model estimates $m_{s,\text{ds}}$ (purple points) vs. $\Phi$. Right: Histogram of residuals $m_s-m_{s,\text{dss}}$.}
        \label{fig:dss_estmag_all}
\end{SCfigure*}
\afterpage{
\begin{figure*}[h!]
    \begin{center}
        \includegraphics[width=17.cm]{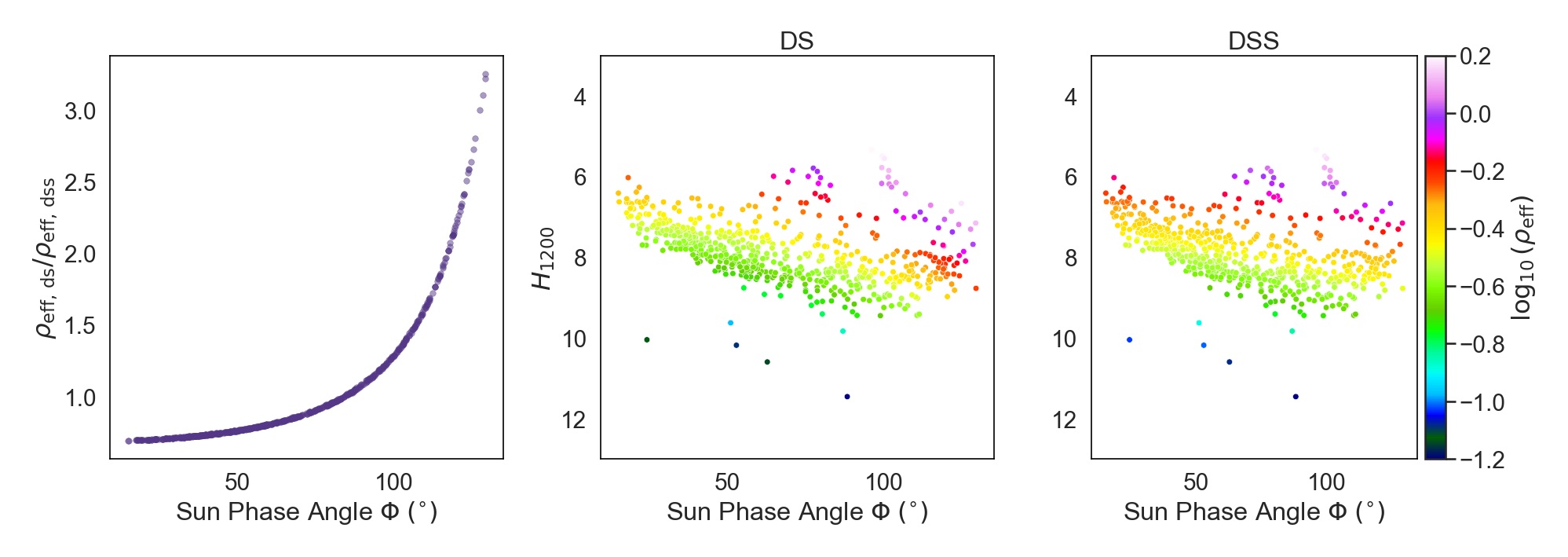}
    \end{center}
    \caption{Effective albedo for the diffuse and half-specular spheres. Left: $\rho_{\text{eff,ds}}/\rho_{\text{eff,dss}}$ vs. $\Phi$. Center and right: $H^{1200}$ vs. $\Phi$, color-coded by $\rho_\text{eff,ds}$ and $\rho_\text{eff,dss}$, respectively. For the same effective albedo, $H^{1200}$ follows the shape for each model (see Appendix \ref{apx:more_figures}).}
    \label{fig:ea_phaseangle}
\end{figure*}
}

We used the differential evolution Markov chain Monte Carlo (DEMC) method to fit the data with the proposed brightness models. This method, which combines the Markov chain Monte Carlo and differential evolution algorithms, makes the search process more efficient by avoiding sampling local minimal approximations. The method adjusts multiple chains in parallel that learn from each other instead of running independently. New chains are accepted or rejected according to the Metropolis-Hastings criterion \citep{Sherri2019}.

A Python routine was created using the \texttt{PyMC} library to fit the DS and DSS models to the stationary magnitudes $m_{s}$, and the RR model to the range-corrected magnitudes $H^{1200}$, using DEMC. All brightness models were computed for all satellites, the on-station, and the off-station subsets. 

In general, to use the \texttt{PyMC} library and DEMC, users must perform the following: assign the prior distributions of both the parameters and the noise of the observations; define the deterministic function to be fit, which depends on the parameters and the observations; define the likelihood of the observations, which in this case corresponds to a normal distribution of the observations around the function being fitted (see Table \ref{tab:demc_parameters}); and, lastly, define the step method and the parameters for iterations. In our case, everything was constant for the three models (DS, DSS, RR). The Metropolis-Hastings criterion (\texttt{Metropolis}, class of \texttt{PyMC}) was used, setting the tuning frequency (\texttt{tune\_interval}) equal to 150 and the initial scale factor for proposal (\texttt{scaling}) equal to 0.01. The iteration parameters were set using the \texttt{sample} function of \texttt{PyMC}. Moreover, following parameters were set: the number of samples to draw (\texttt{draws}) equal to 25\,000, number of iterations to tune (\texttt{tune}) equal to 25\,000, number of chains to sample (\texttt{chains}) equal to 10, step function or collection of functions (\texttt{step}) equal to \texttt{Metropolis}, and random seed(s) used by the sampling steps (\texttt{random\_seed}) equal to 42.

\section{Results}\label{sec:results}

\begin{figure*}[t!]
    \begin{center}
        \includegraphics[width=19.cm]{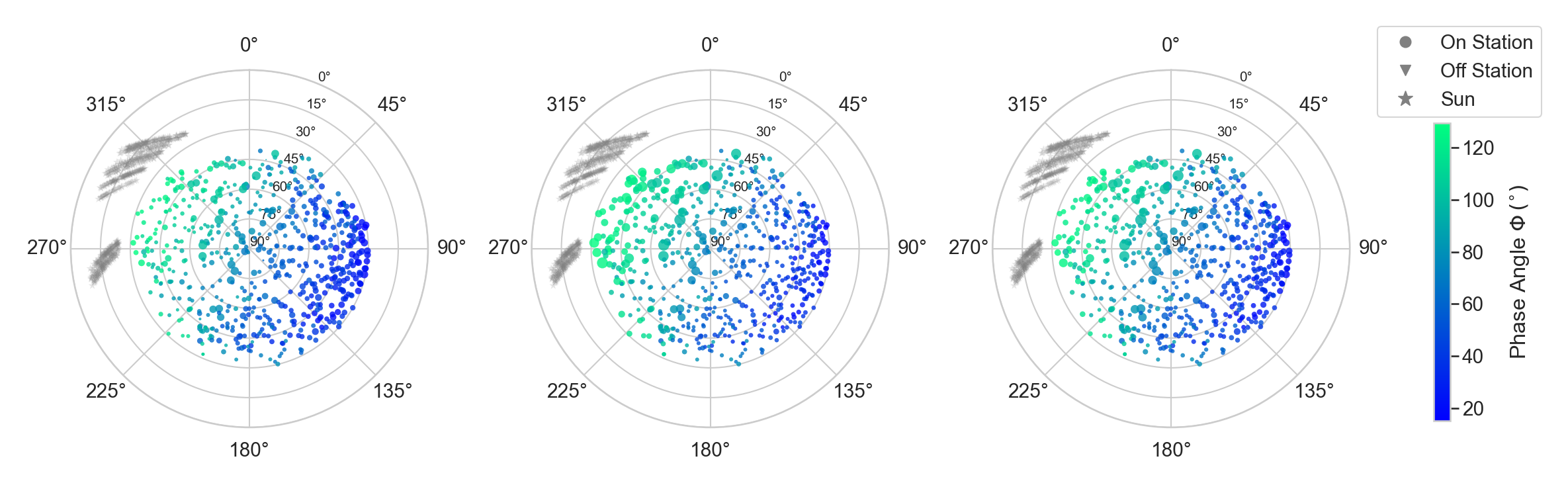}
        \caption{Horizontal coordinates of the ONEWEB LEOsats and the Sun projected on a polar plane. North and east are at $0^\circ$ and $90^\circ$, respectively. Both visible and non-visible sky are projected onto the same plane. The Sun is always below the horizon. Left panel: LEOsats range-corrected magnitude $H^{1200}$ distribution. The larger the symbol, the brighter the satellite. Center panel: LEOsats diffuse sphere effective albedo distribution. The larger the symbol, the larger the effective albedo $\rho_{\text{eff,ds}}$. Right panel: Same as center panel, but for the half-specular sphere effective albedo $\rho_{\text{eff,dss}}$.}
        \label{fig:sky_plot}
    \end{center}
\end{figure*}

\subsection{Lambertian diffuse spherical model}

The DS model in equations (\ref{eq:lambertian_model}) and (\ref{eq:f0_ds}) describes the stationary magnitude $m_{s}$ as a function of the Sun phase angle $\Phi$ and fixed parameters $\zeta$ and $p$. The DEMC algorithm finds the values of $\zeta$ and $p$ for which the DS model reproduces best the magnitudes $m_{s}$. The input data are $m_{s}, \Phi$, and $r$, and the resulting parameters can be seen in Table \ref{tab:all_models}, for all satellites, on-, and off-station samples. Figure \ref{fig:ds_estmag_all} shows the observed stationary magnitudes and the ones computed by the model, as well as the residuals between them, for all satellites. The same parameters results (within $10^{-3}$) are obtained when the fitting is performed using scaled magnitudes $H^{1200}$. Parameters results for the V filter and R filter samples can be seen in Tables \ref{tab:v_filter_results_models} and \ref{tab:r_filter_results_models}, respectively, in Appendix \ref{apx:more_results_table}.

\subsection{Lambertian diffuse specular spherical model}

Likewise, in the previous section, the DSS model in equations (\ref{eq:lambertian_model}) and (\ref{eq:f0_dss}) describes the stationary magnitude $m_{s}$ as a function of $\Phi$ and fixed parameters $\zeta$ and $\beta$. The DEMC algorithm finds the values of $\zeta$ and $\beta$ for which the DSS model reproduces best the magnitudes $m_{s}$. The input data are the same than in the DS model, and the resulting parameters can be seen in Table \ref{tab:all_models}, for all satellites, on-, and off-station samples. 
Figure \ref{fig:dss_estmag_all} shows the observed stationary magnitudes, the model-computed magnitudes, and the residuals between them, for all satellites. Again, the same parameters results (within $10^{-3}$) are obtained when the fitting is performed using scaled magnitudes $H^{1200}$. Parameters results for the V filter and R filter samples can be seen in Tables \ref{tab:v_filter_results_models} and \ref{tab:r_filter_results_models}, respectively, in Appendix \ref{apx:more_results_table}.

\subsection{DS versus DSS results}
For each stationary magnitude observed and plotted in the left panels of Figures \ref{fig:ds_estmag_all} and \ref{fig:dss_estmag_all}, the results from the DS and DSS models, respectively, are very similar. Both plots look almost the same; however, point by point, they differ typically within $\pm 0.051$ magnitudes. 

\begin{figure*}[t!]
    \includegraphics[ trim=0.5cm 0.5cm 1cm 0cm,width=7in]{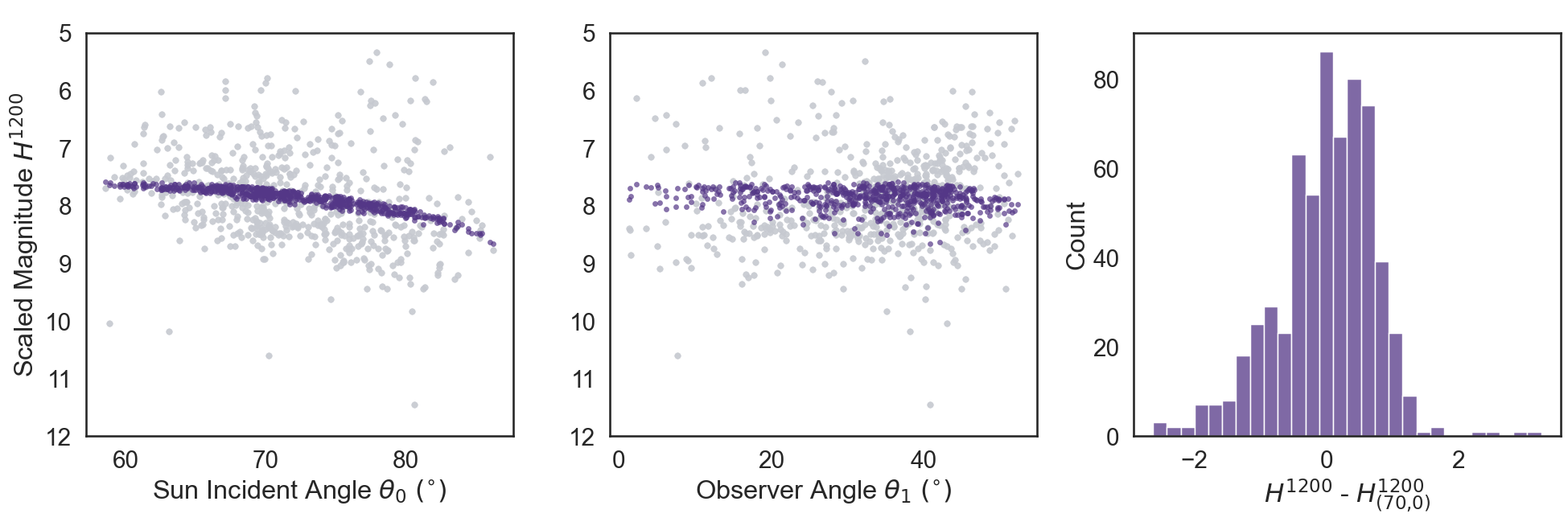}
    \caption{Results of the RR model applied to all the ONEWEB LEOsats of this study.
    Left and center: Observed $H^{1200}$ (gray points) and DSS model estimates $H^{1200}_\text{rr}$ (purple points) vs. $\theta_0\,$ and $\,\theta_1$, respectively. Right: Histogram of residuals $H^{1200}-H^{1200}_\text{rr}$.}
    \label{fig:dss_estmag_rr}
\end{figure*}

\subsection{Effective albedo for the DS and DSS models}\label{subsec_effalbedo}

Some algebra proves that the ratio $\rho_{\text{eff,ds}}/\rho_{\text{eff,dss}}$ follows a defined curve below 1 for $\Phi\sim 84^\circ$ and greater than 1 for larger values of $\Phi$. This curve is plotted in the left panel of Figure \ref{fig:ea_phaseangle}. In other words, each observation can be associated to either model's effective albedo but a diffuse sphere must be less reflective for $\Phi\lesssim 84^\circ$ and more reflective for $\Phi\gtrsim 84^\circ$ than the half-specular one. What seems at first counterintuitive has an explanation. An observer receives flux from the whole visible illuminated surface of a diffuse sphere because all parts of the illuminated surface reflect in all directions. For the half-specular version, the light it reflects - which for any differential surface area is more than its diffuse counterpart - will reach the observer only if it is properly aligned. The total flux received, of course, also depends on the orientation, as parameterized by $\Phi$. When we observe a fully illuminated satellite face-on (i.e., at full phase or small $\Phi$), the diffuse sphere does not need to be so reflective to explain the observed brightness. When we observe only a small illuminated portion of the satellite (i.e., close to new phase or large $\Phi$), the diffuse sphere must be more reflective to fit the observed magnitude as compared to the half-specular version, for which a smaller illuminated area but precisely oriented reflects as much light.

From equations (\ref{eq:effective_albedo_ds}) and (\ref{eq:effective_albedo_dss}), we computed for every single observation its corresponding effective albedo, one for each model. Figure \ref{fig:ea_phaseangle} shows the distribution of $H^{1200}$ versus $\Phi$, color-coded by the effective albedo $\rho_\text{eff,ds}$ (middle panel) and $\rho_\text{eff,dss}$ (right panel), respectively. It is easy to notice that the observations with the same effective albedo follow the expected shape for each model (Figures \ref{fig_apx:ds_model_by_p_A_rho} and \ref{fig_apx:dss_model_by_beta_A_rho} in Appendix \ref{apx:more_figures}.). Also, as expected, at each Sun phase angle $\Phi$, larger effective albedo yield brighter magnitudes. The overall observed trend for most of the range-scale magnitudes $H^{1200}$ in Figure \ref{fig:ea_phaseangle} is as follows:
\begin{itemize}
    \item Small $\Phi$ (large $\gamma$): The satellite is close to full phase; solar panels point to the Sun and also roughly toward the observer. Therefore the satellite is brighter. We observe a lot of reflection from the solar panels, and they have a larger area than the bus. 
    \item Large $\Phi$ (small $\gamma$): The satellite is close to new phase; solar panels point to the Sun and away from the observer. If it is close enough to the Sun, the satellite is not actually visible. But, when the Sun is below the horizon, and the satellite is at an angular distance $\gamma$ large enough from the Sun, its bus side toward Earth and its antennas (also pointing toward Earth) may act as a mirror, making the satellite visible and bright. This would explain the small increase in brightness observed for $\Phi\gtrsim 100^\circ$. 
\end{itemize} 

This is readily seen in the polar plots of the horizontal coordinates of the satellites and the Sun shown in Figure \ref{fig:sky_plot}. The satellites' symbols are colored according to the corresponding Sun phase angle $\Phi$. The left panel shows how the satellites' range-corrected magnitude is generally brighter when they are opposite the Sun (full phase) and fainter with higher dispersion for the rest of the positions. This happens for both off- and on-station satellites. The increased brightness at full phase is most probably due to specular reflection by the LEOsats solar panels oriented consistently toward the Sun. The center and right panels show the effective albedos, indicated by the symbol size, for the diffuse and the half-specular models, respectively. At full phase or small $\Phi$, the diffuse sphere yields slightly lower effective albedos than the half-specular one. At new phase or large $\Phi$, the opposite occurs, the diffuse sphere yields in average higher effective albedos than the half-specular one. Overall, the diffuse sphere effective albedo shows more visible mean changes in horizontal coordinates than the half-specular one. As proposed by \citet{Krantz2023}, the diffuse sphere effective albedo is a measure of the reflection's specularity, thereby possibly tracing the different attitudes of the ONEWEB satellites. On the other hand, the half-specular sphere effective albedo appears less structured in horizontal coordinates. It can either be closer to the real mean physical albedo of the ONEWEB satellites or the half-specular component absorbs or partially accounts for the specularity changes. 

In Appendix \ref{apx:more_figures}, Figure \ref{fig_apx:delta_az}, we show additional plots with respect to the azimuth coordinates, for comparison with work such as \citet{Mallama2021sep, Mallama2021nov} and \citet{mroz2022}. The observed trends of $H^{1200}$ with $\Delta Az$ are explained by the strong correlations in specific intervals of $\Delta Az$ with $\Phi$. In our data, we find that the range-corrected magnitude follows a tighter zig-zag pattern with $\Delta Az$ for the 2021 observations (October to November, the Sun at sunset to the south), while the 2022 observations (April to September, the Sun at sunset to the north) follow a smoother more disperse oscillation. These two groups correspond to the two clusters of gray symbols in Figure \ref{fig:sky_plot}. No difference is found in the plot $H^{1200}$ versus $\Phi$, although at the low end of $\Phi$, the 2021 observations are, on average, slightly brighter, up to about half a magnitude for $\Phi<25^\circ$ (see Figure \ref{fig_apx:delta_az}). More extended observations are needed to discard a bias effect caused by the Sun's azimuth interval of the 2021 observations being half the size of that of the 2022 observations. Another possibility is a seasonal and/or geographical variation of the Earth's albedo playing a role, as discussed in Section \ref{sec:discussion}.

We do notice a few observations in Figure \ref{fig:ea_phaseangle} that deviate from what we just described. The faintest observations for which the effective albedo is the lowest, and a nonnegligible number of the brightest observations and largest effective albedos (at $\Phi\sim 80^\circ$ and $\Phi\sim 100^\circ$). The latter appear in Figure \ref{fig:sky_plot} as the largest size symbols, more or less following a line from direction NNE to WSW. We checked these subsamples in detail but find nothing remarkable that could explain their extreme values.

\subsection{Relative reflective model - RR}
For the RR model, equation (\ref{eq:RR}) describes the range-corrected magnitude $H^{1200}$ as a function of $(\theta_0,\theta_1)$ and the fixed parameters $k$, and $H^{1 200}_{(70,0)}$ which measures the constant brightness of the satellites at 1\,200 km range and fixed orientation $(70^\circ,0^\circ)$. The DEMC algorithm finds the values of $k$ and $H^{1 200}_{(70,0)}$ for which the RR model reproduces best the magnitudes $H^{1200}$. The input data are $H^{1200}$ and $(\theta_0,\theta_1)$, and the resulting parameters can be seen in Table \ref{tab:all_models}, for all satellites, on-, and off-station samples. Figure \ref{fig:dss_estmag_rr} shows the observed range-corrected magnitudes the model-computed magnitudes, the residuals between them, for all satellites. Parameters results for the V filter and R filter samples can be seen in Tables \ref{tab:v_filter_results_models} and \ref{tab:r_filter_results_models}, respectively, in Appendix \ref{apx:more_results_table}.

\section{Discussion and conclusion}\label{sec:discussion}

In all models, we find that the residuals have a dispersion of about 0.6 - 0.7 magnitudes when having a large sample of data points ($\sigma_\text{demc}$ in Table \ref{tab:demc_parameters}). That amount cannot be explained by the input data errors. The observed $m_s$ are affected only by photometric errors, which are typically about 0.2 magnitudes. Errors in range $r$ are negligible, since the TLE-SGP4-ranges have estimated errors below 1 km (see e.g., Figures 3 to 5 in \citet{Vallado2012} and Figure 4 in \citet{geul2017}).

Therefore, although all the brightness models used can fit the mean value of the observed stationary magnitudes, they still lack other details that account for the rest of the observed residuals. More importantly, we noticed that the neither model reproduces the overall shape of $m_s$ versus $\Phi$ (DS, DSS) or $H^{1200}$ versus $(\theta_0,\theta_1)$ (RR) plots. The fact is that both Lambertian models yield fainter magnitudes as $\Phi$ increases, but we observe a slight increase in $H^{1200}$ for $\Phi\gtrsim 100^\circ$ (see Figure \ref{fig:ea_phaseangle}). 

We therefore conclude that a spherical shape is a first approach to modeling the observed brightness of the ONEWEB LEOsats. However, there is a need for improvement to reproduce the brightness observations within the limits of their errors and the observed trend that seems to exist for larger $\Phi$. In this work, we proved that spherical and RR models, whose magnitude modulation depends fundamentally on the relative position Sun-satellite-observer as parameterized by $\Phi$ (or $(\theta_0,\theta_1)$) but are independent of the orientation of the body itself (attitude) cannot explain the observations well enough. This means that more information is needed regarding the specific orientation (and size and albedo) of the various surfaces, that are actually reflecting the Sun's light onto Earth, with respect to the Sun and the observer. Since we know the reflecting surfaces of the ONEWEB satellites are close to flat (including the antennas), a flat reflecting surface model is being considered for future work by the authors. \citet{Mccue_1971} cites flat surface brightness models from \citet{Liemohn1968} (specular) and  \citet{Giese1963} (diffuse). In \citet{Maley1987}, the specular flat plate model is cited as the analytical justification for the observed flashes or outburst of brightness for several Cosmos satellites. Those presumed to be tumbling could generate random flashes. ONEWEB satellites examined by this work do not, however, show such behavior.  

Another factor that has not been considered so far is Earth's albedo. Our planet reflects sunlight to space and some of it reaches the satellites, LEOsats being closer are therefore more affected. In other words, the flux received by the satellite is higher than $F_\odot$.  If we apply the formalism explained in Appendix \ref{apx:lambertian} to Earth as the reflecting sphere, i.e., we model Earth as a Lambertian sphere and express the flux from Earth that reaches the satellite $F_\oplus=\kappa F_\odot$ with $\kappa>0$, then $F=(1+\kappa)F_\odot$ and the new magnitude will be brighter by $-2.5\log(1+\kappa)$. More specifically
\begin{equation}
\kappa=A_{B,\oplus}\;\pi\; F_{0,\oplus}(z_{\odot,\text{sat}})\left(\frac{R_\odot}{H_\text{sat}}\right)^2 \; , \label{eq:kappa}    
\end{equation}
with $A_{B,\oplus}$ being the Earth's Bond albedo and $F_{0,\oplus}$ the corresponding phase function in terms of the phase angle which in this case is $z_{\odot,\text{sat}}$. The last term in (\ref{eq:kappa}) between $\sim$25 and $\sim$120 for our sample of satellites. If Earth were a Lambertian diffuse sphere then its phase function $F_{0,\oplus}$ would be the DS model in equation (\ref{eq:f0_ds}). For a fixed $H_\text{sat}$ (e.g., the on-station sample), such $F_{0,\oplus}$ decreases with $z_{\odot,\text{sat}}$, similar to the associated flux on the satellite. Earth flux is greater when the Sun is closer to the satellite's zenith (smaller $z_{\odot,\text{sat}}$) and the Earth looks fuller from it.

In our sample, there is anticorrelation between $z_{\odot,\text{sat}}$ and $\Phi$, although with a high dispersion: $125^\circ\gtrsim z_{\odot,\text{sat}}\gtrsim 90^\circ$ when $15^\circ\lesssim\Phi\lesssim 130^\circ$. The observed mean increase in $H^{1200}$ versus $\Phi$ for $\Phi\gtrsim 100^\circ\; (z_{\odot,\text{sat}}\lesssim 110^\circ$) could be explained if for these data points, Earth's albedo becomes important enough to increase the range-corrected magnitudes. In the absence of earthshine $H^{1200}$ should keep decreasing. Work studying earthshine from observations of the Moon at various phases have found that the Earth’s phase function $F_{0,\oplus}$ is observed to be roughly Lambertian around crescent Moon phases (or gibbous Earth, as seen from the Moon) \citep{Qiu2003, Robinson2010, Goode2021}. In such a case, the Earth's apparent albedo is roughly comparable to the Earth's visible light albedo ($\sim$0.3) \citep{Goode2021}. It is important to note that Earth appears Lambertian at gibbous phases not because it truly is Lambertian, but because thick cloud cover and geometric averaging suppress anisotropic scattering, making the disk-integrated reflectance mimic that of a diffuse sphere. On the other hand, crescent and full Earth disk-integrated brightnesses deviate significantly from the Lambertian model. Forward scattering by clouds and aerosols, as well as ocean specular glint can dominate total flux for crescent phases, while coherent backscatter and shadow hiding (especially over clouds and ice) are important in the full phase \citep{Robinson2010, Goode2021}. Moreover, the long-term monitoring of earthshine (e.g., \citet{Goode2021}) has found temporal variation of the Earth albedo within the Lambertian behavior tied to the seasons and climate cycles. 

Another important factor is that this approximation holds for distant observers, the Moon is far enough from Earth but LEOsats are not. For a distant observer all the surface elements of the Earth's sphere can be assumed to be at the same phase angle, while for a LEOsat, different surface elements on Earth have measurably different phase angles. In any case, a Lambertian diffuse and specular Earth may be considered as a first approach to model the Earth albedo effect on the observed brightness of LEOsats.

\bibliographystyle{aa}
\bibliography{biblio}

\section{Data availability}
The full content of Table \ref{tab:tle_transformation} can be found at \url{https://doi.org/10.5281/zenodo.18963398}.

\begin{acknowledgements}
    Author MRC acknowledges the Universidad de Atacama PhD Scholarship; and the Air Force Office of Scientific Research  (AFORS) grant \# FA9550-22-1-0292.
\end{acknowledgements}

\begin{appendix}
\nolinenumbers
\onecolumn

\section{Observations and results tables for V filter and R filter samples}\label{apx:more_results_table}

\begin{table}[!h]
    \begin{center}
    
    \caption{Observational information for each sample studied.}
    \setlength{\tabcolsep}{5pt} 
    \begin{tabular}{cc>{\columncolor[HTML]{EFEFEF}}cc>{\columncolor[HTML]{EFEFEF}}c|c>{\columncolor[HTML]{EFEFEF}}cc>{\columncolor[HTML]{EFEFEF}}c}
    \cline{2-9}
         & \multicolumn{4}{c|}{On-Station}                                         & \multicolumn{4}{c}{Off-Station}                                \\ \cline{2-9} 
         & Number of & Stationary  & Altitude  & Range & Number of & Stationary    &  Altitude  & Range     \\
         & Satellites &  Magnitude & (km) & (km)  & Satellites & Magnitude  & (km) & (km)  \\ \hline
V filter & 410         & 8.44 $\pm$ 0.6  & 1211.3 $\pm$ 9.3  & 1547.0 $\pm$ 204.2  & 40  & 7.15±0.9 & 684.7 $\pm$ 110.8 & 919.4 $\pm$ 184.3 \\ \hline
R filter & 158         & 8.46  $\pm$ 0.9 & 1212.7 $\pm$ 5.3 & 1494.7  $\pm$ 172.4 & 17 & 6.73±1.6 & 686.8 $\pm$ 91.1   & 867.8 $\pm$ 156.4 \\ \hline
    \end{tabular}
    \label{tab:stad_satellite_altitud}
    \tablefoot{Average and standard deviation of the stationary magnitude, altitude, and range, are shown per orbit type and filter.}
    \end{center}
\end{table}

\renewcommand{\arraystretch}{1.7} 
\begin{table*}[!h]
\begin{center}
    \caption{DEMC-fitted parameters and errors of the DS, DSS, and RR brightness models, for the V filter samples.}
    \begin{tabular}{ccc>{\columncolor[HTML]{EFEFEF}}cc>{\columncolor[HTML]{EFEFEF}}c}
                                &                            && V filter     & V filter On Station        & V filter Off Station      \\ \hline 
                                & $p$                        && 0.402\;$\pm$\;0.041 & 0.445\;$\pm$\;0.047 & 0.364\;$\pm$\;0.083 \\ 
                                & $\zeta$                    &(m$^2$)& 0.127\;$\pm$\;0.005 & 0.125\;$\pm$\;0.006 & 0.142\;$\pm$\;0.013 \\
    \multirow{-3}{*}{DS Model}  & $\sigma_{\text{demc,ds}}$     &(mag)& 0.679\;$\pm$\;0.020 & 0.632\;$\pm$\;0.022 & 0.775\;$\pm$\;0.044 \\ \hline
                                & $\beta$                    && 0.215\;$\pm$\;0.006 & 0.203\;$\pm$\;0.006 & 0.249\;$\pm$\;0.014 \\
                                & $\zeta$                    &(m$^2$)& 0.430\;$\pm$\;0.040 & 0.465\;$\pm$\;0.046 & 0.395\;$\pm$\;0.083 \\ 
    \multirow{-3}{*}{DSS Model} & $\sigma_{\text{demc,dss}}$     &(mag)& 0.663\;$\pm$\;0.020 & 0.614\;$\pm$\;0.021 & 0.762\;$\pm$\;0.043 \\ \hline
                                & $k$                        && 0.489\;$\pm$\;0.075 & 0.514\;$\pm$\;0.082 & 0.243\;$\pm$\;0.141 \\
                                & $H^{1200}_{(70,0)}$        &(mag)& 7.691\;$\pm$\;0.040 & 7.760\;$\pm$\;0.042 & 7.416\;$\pm$\;0.091 \\
    \multirow{-3}{*}{RR Model}  & $\sigma_{\text{demc,rr}}$     &(mag)& 0.704\;$\pm$\;0.021 & 0.669\;$\pm$\;0.023 & 0.768\;$\pm$\;0.043 \\ \hline
    \end{tabular}
    \label{tab:v_filter_results_models}
\end{center}

\end{table*}

\renewcommand{\arraystretch}{1.7} 
\begin{table*}[!h]
\begin{center}
    \caption{DEMC-fitted parameters and errors of the DS, DSS, and RR brightness models, for the R filter sample.}
    \begin{tabular}{ccc>{\columncolor[HTML]{EFEFEF}}cc>{\columncolor[HTML]{EFEFEF}}c}
                                &                            && R filter     & R filter On Station       & R filter Off Station      \\ \hline 
                                & $p$                        && 0.215\;$\pm$\;0.147 & 0.350\;$\pm$\;0.165 & 0.458\;$\pm$\;0.277 \\ 
                                & $\zeta$                    &(m$^2$)& 0.131\;$\pm$\;0.031 & 0.127\;$\pm$\;0.029 & 0.346\;$\pm$\;0.186 \\
    \multirow{-3}{*}{DS Model}  & $\sigma_{\text{demc,ds}}$     &(mag)& 1.130\;$\pm$\;0.107 & 0.857\;$\pm$\;0.099 & 1.581\;$\pm$\;0.261 \\ \hline
                                & $\beta$                    && 0.270\;$\pm$\;0.037 & 0.227\;$\pm$\;0.028 & 0.524\;$\pm$\;0.212 \\
                                & $\zeta$                    &(m$^2$)& 0.218\;$\pm$\;0.151 & 0.368\;$\pm$\;0.167 & 0.402\;$\pm$\;0.268 \\ 
    \multirow{-3}{*}{DSS Model} & $\sigma_{\text{demc,dss}}$     &(mag)& 1.124\;$\pm$\;0.106 & 0.846\;$\pm$\;0.098 & 1.566\;$\pm$\;0.260 \\ \hline
                                & $k$                        && 0.610\;$\pm$\;0.240 & 0.546\;$\pm$\;0.237 & 0.488\;$\pm$\;0.283 \\
                                & $H^{1200}_{(70,0)}$        &(mag)& 7.577\;$\pm$\;0.221 & 7.728\;$\pm$\;0.196 & 7.014\;$\pm$\;0.445 \\
    \multirow{-3}{*}{RR Model}  & $\sigma_{\text{demc,rr}}$     &(mag)& 1.120\;$\pm$\;0.106 & 0.863\;$\pm$\;0.099 & 1.502\;$\pm$\;0.248 \\ \hline
    \end{tabular}
    \label{tab:r_filter_results_models}
\end{center}

\end{table*}

\FloatBarrier

\twocolumn
\section{Magnitudes and phase function}\label{apx:lambertian}
From the definitions of magnitudes and fluxes, we want to compare the flux we receive from a sphere $F_\text{sph}$ to the flux we receive from the Sun, then
\[
m_\text{sph}-m_\odot = -2.5\log\left(\frac{F_\text{sph}}{F_\odot} \right)\,.
\]
We start with the definitions of geometric albedo $p$ and phase function $F(\Phi)$, as in \citet[Section 5.5]{Shepard2017}:
\begin{eqnarray*}
  p &=& \frac{I(0)}{F\, R^2} \;,\\
  F(\Phi) &=& \frac{I(\Phi)}{I(0)} \;,
\end{eqnarray*}
where $F$ is the flux received by the sphere, $R$ is its radius, $I$ is the radiant intensity from the sphere and $\Phi$ is the phase angle of that flux. As noted by \citet[Section 5.5.2]{Shepard2017}, the geometric albedo compares the reflectance of the sphere to that of a 
flat disk made of Lambertian material with the same radius $R$, both seen face-on. $I(\Phi)$ is the amount of energy we receive from (reflected by) the sphere per unit time and unit solid angle. Therefore:
\[
F_\text{sph}(\Phi)=\frac{I(\Phi)}{r^2}=\frac{I(0)F(\Phi)}{r^2}=\frac{p\, F\, R^2 F(\Phi)}{r^2} \, .
\]
From \citet[Section 5.5.4]{Shepard2017}, we have the definitions of phase integral $q$ and Bond (or spherical) albedo $A_B$:
\begin{eqnarray*}
    q &=& \frac{\int_0^{4\pi}F(\Phi)d\Omega}{\pi} \;,\\
    A_B &=& p \cdot q \;, 
\end{eqnarray*}
from which we obtain that $\displaystyle p=\frac{A_B \pi}{\int_0^{4\pi} F(\Phi)d\Omega}$. Substituting in the equation for the sphere flux above, we have
\begin{eqnarray*}
F_\text{sph}(\Phi) &=& \frac{A_B \pi}{\int_0^{4\pi} F(\Phi)d\Omega}\frac{F\, R^2 F(\Phi)}{r^2} \\
&=& \frac{A_B\, F\, \pi R^2}{r^2} \frac{F(\Phi)}{\int_0^{4\pi} F(\Phi) d\Omega } \\
&=& \frac{A_B\, F\, \pi R^2}{r^2} F_0(\Phi)\, .
\end{eqnarray*}
$F_0(\Phi)$ is referred to by \citet{Williams1966} as the phase function normalized by the integral over a solid angle of $4\pi$ stereoradians. 
Substituting $F=F_\odot$, i.e., the sphere is illuminated by the Sun at 1 AU, we obtain
\begin{eqnarray*}
\frac{F_\text{sph}(\Phi)}{F_\odot} &=& \frac{A_B\;\pi R^2\; F_0(\Phi)}{r^2} \Longrightarrow  \\
m_\text{sph} - m_\odot &=& - 2.5 \log_{10}\left(\frac{F_\text{sph}(\Phi)}{F_\odot} \right) \Longrightarrow  \\
m_\text{sph} &=& m_\odot - 2.5 \log_{10}\left[A_B\; \pi R^2\; F_0(\Phi)\right]+5\log_{10}(r) \, .
\end{eqnarray*}
In equation (\ref{eq:lambertian_model}): $m_s=m_\text{sph}$, $\;\rho=A_B$, $\;A=\pi R^2$ is the cross section of the sphere, and $\zeta= \rho\, A$. For a Lambertian sphere, the Bond Albedo is equal to the albedo \citep[equation 5.71]{Shepard2017}. Keep in mind that the units of $A$ have to be compatible with the units of $r$.
In other words, if the sphere cross section is given in
square meters, the range has to be given in meters \citep{karttunen2003fundamental}. 

\section{Additional figures}\label{apx:more_figures}
\begin{figure}[!h]
    \includegraphics[width=3in]{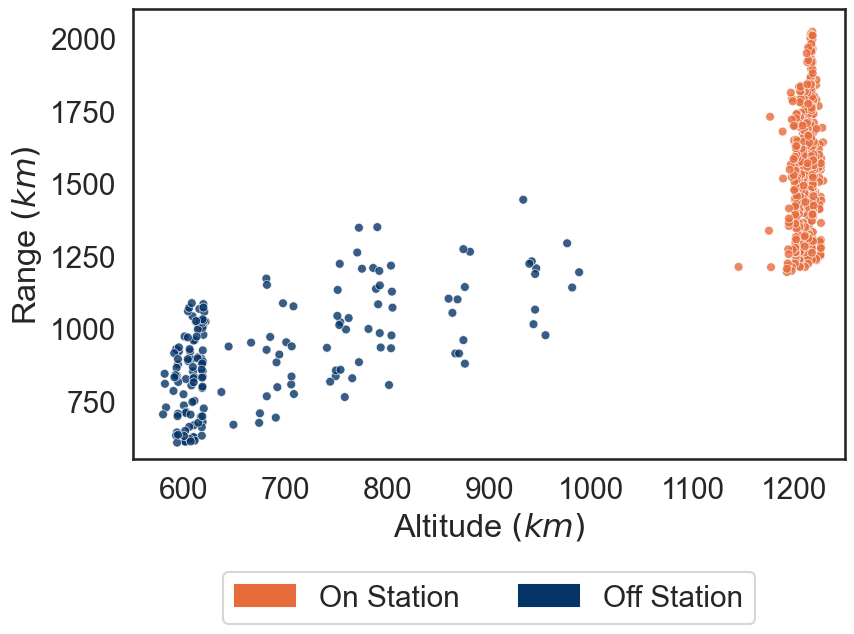}
    \caption{ONEWEB LEOsats of this investigation, separated by their on-station (orange) and off-station (blue) orbital stages based on their altitude. The vertical axis shows their range, i.e., distance between the observer and the satellite. Off- and on-station satellites have different altitudes, but they may have similar ranges.}
    \label{fig_apx:satellite_altitude_distribution}
\end{figure}

\begin{figure}[!h]
    \includegraphics[width=3.5in]{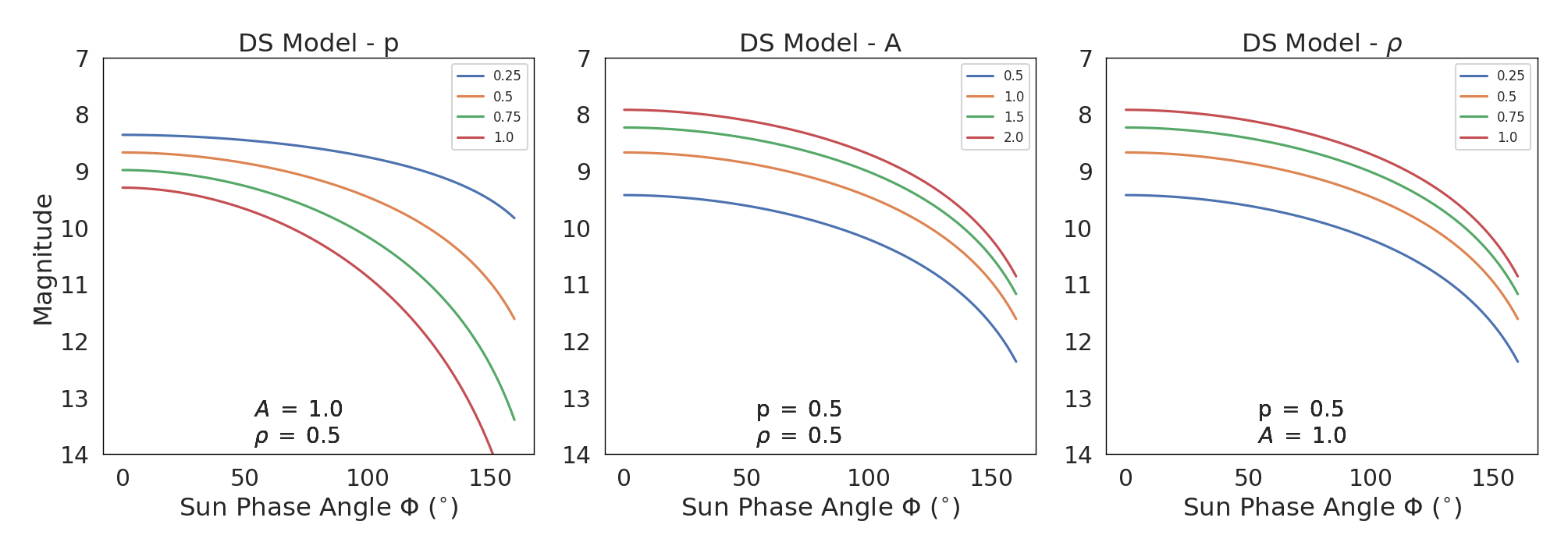}
    \caption{Plots of $m_s$ for the DS model, considering different values of $p$, $\rho$, and $A$, from left to right panels, respectively, as indicated by the legends. All plots consider $r=1\,200$ km. Left: $A=1$ and $\rho=0.5$. Middle: $p=\rho=0.5$. Right: $p=0.5$ and $A=1$.}
    \label{fig_apx:ds_model_by_p_A_rho}
\end{figure}

\begin{figure}[!h]
    \includegraphics[width=3.5in]{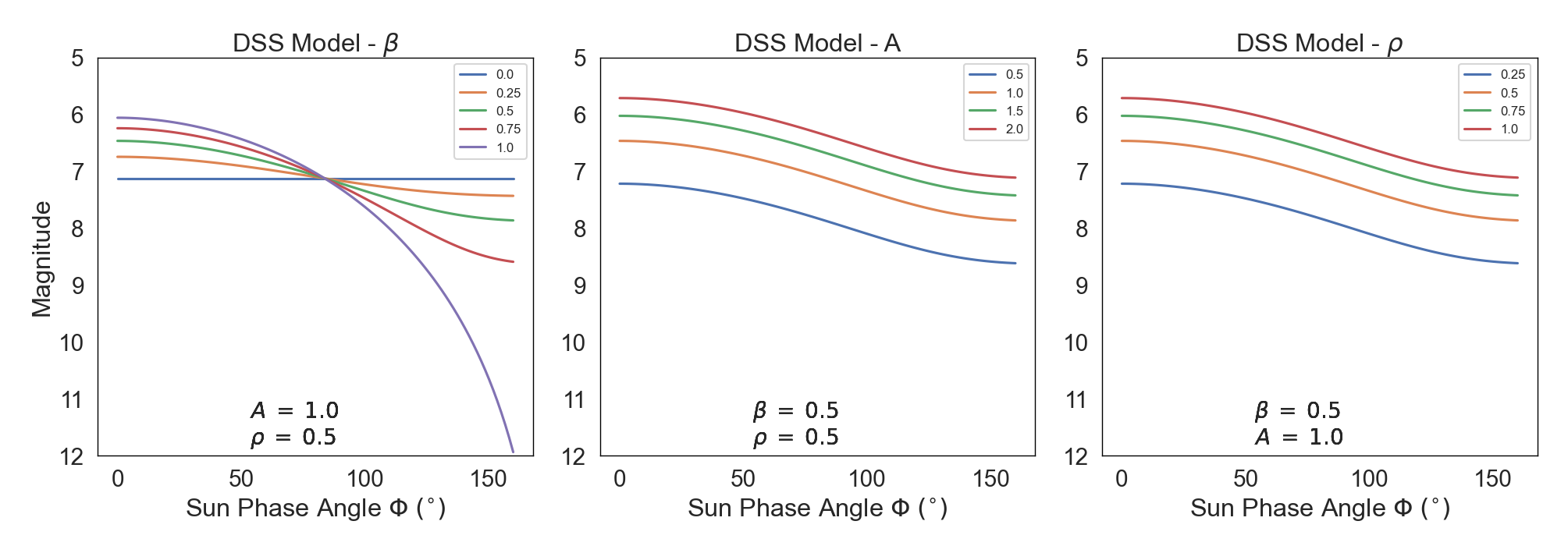}
    \caption{Plots of $m_s$ for the DSS model, considering different values of $\beta$, $\rho$, and $A$, from left to right panels, respectively, as indicated by the legends. All plots consider $r=1\,200$ km. Left: $A=1$ and $\rho=0.5$. Middle: $\beta=\rho=0.5$. Right: $\beta=0.5$ and $A=1$.}
    \label{fig_apx:dss_model_by_beta_A_rho}
\end{figure}

Figures \ref{fig_apx:satellite_altitude_distribution}, \ref{fig_apx:ds_model_by_p_A_rho} and \ref{fig_apx:dss_model_by_beta_A_rho} contain complementary information.

\begin{figure}[ht!]
    \begin{center}
        \includegraphics[ trim=0.5cm 0.5cm 1cm 0cm,width=3.5in]{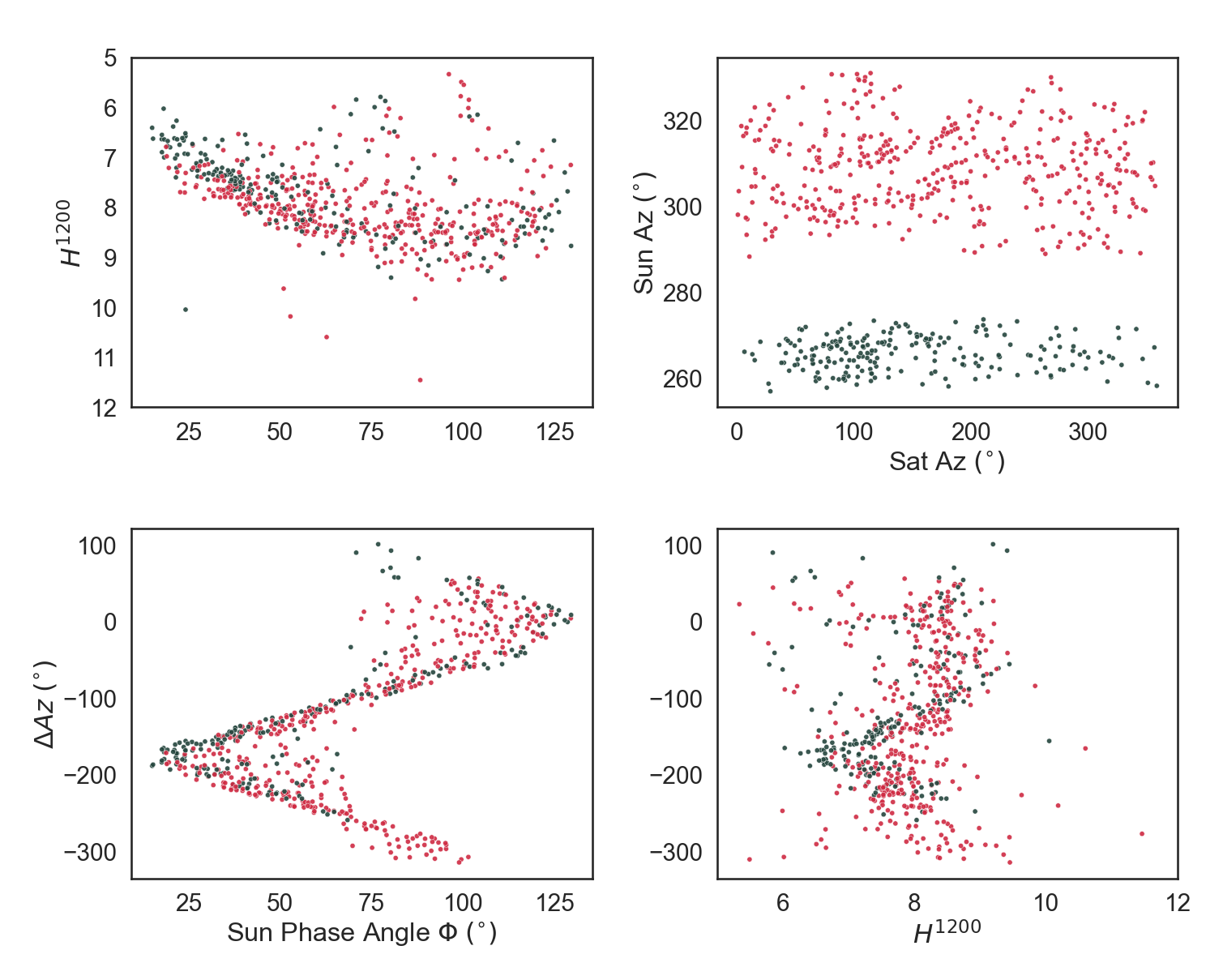}
        \caption{Variation of $H^{1200}$ with azimuth. In all figures, points are color-coded by the azimuth of the Sun: north is red, south is green. Upper left: $H^{1200}$ vs. $\Phi$, as in Figure \ref{fig:ea_phaseangle}, for reference. Lower left: $\Delta Az$ = $Az_\text{sat}-Az_\odot$ vs. $\Phi$. Lower right: $\Delta Az$ vs. $H^{1200}$. Upper right: $Az_\odot$ vs. $Az_\text{sat}$.}
        \label{fig_apx:delta_az}
    \end{center}
\end{figure}

\FloatBarrier

\section{Glossary of terms}\label{apx:glossary}

\begin{itemize}
    \item $m_{s}$ = stationary magnitude.
    \item $m_{0}$ = zero-point magnitude of the stars.
    \item $m_{\text{ins}}$ = instrumental aperture magnitude.
    \item $\phi$ = angular velocity satellite in the sky.
    \item $L$ = length of the line left by the LEOsat.
    \item $t_{\text{exp}}$ = exposure time.
    \item $m_{\odot}$ = Sun's apparent magnitude.
    \item $H^{1200}$ = scaled magnitude, range-corrected magnitude.
    \item DS = Diffuse Spherical (model).
    \item DSS = Diffuse Specular Spherical (model).
    \item RR = Relative Reflectance (model).
    \item $\mathcal{R}$ =  ratio of fluxes in RR model.
    \item $F_{0}$ = phase function.
    \item $\zeta$ = effective area of the sphere reflecting light.
    \item $r$ = range of the satellite (satellite-observer distance).
    \item $\rho$ = albedo.
    \item $\rho_{\text{eff,ds}}$ = effective albedo of DS model.
    \item $\rho_{\text{eff,dss}}$ = effective albedo of DSS model.
    \item $A=\pi R^2$ = sphere cross-area.
    \item $\beta$ = mixing coefficient of diffuse-specular reflection.
    \item $F_{\theta_{0}, \theta_{1}}(r)$ = flux coming from direction $\theta_{0}, \theta_{1}$ at a distance $r$.
    \item $k$ = edge darkening parameter.
    \item $\Phi$ = Sun phase angle.
    \item $\theta_{0}$ = Sun incidence angle or solar incidence angle.
    \item $\theta_{1}$ = observer angle.
    \item $\gamma$ = elongation angle.
    \item $H_{(\theta_{0}, \theta_{1})}^{1200}$ = scaled magnitude in orientation $(\theta_{0}, \theta_{1})$.
    \item $H_{(70,0)}^{1200}$ = scaled magnitude in orientation $(70^{\circ}, 0^{\circ})$.
    \item $\Delta$Geog = angular distance between observer's longitude and latitude and the satellite's longitude and latitude.
    \item $\Delta Az = Az_\text{sat}-Az_\odot$ = azimuth of the satellite minus the azimuth of the Sun.
    \item $z_{\odot,\text{obs}}$ = Sun's zenith distance for the observer.
    \item $z_{\odot,\text{sat}}$ = Sun's zenith distance for the satellite.
    \item $O_{\oplus}$ = Sun's Position.
    \item $O_{\text{sat}}$ = satellite's position.
    \item $O_{\text{obs}}$ = observer's position.
    \item $R_{\oplus}$ = Earth's radius.
    \item $H_{\text{sat}}$ = perpendicular height above Earth of the satellite.
    \item $\sigma_{\text{tle}}$ = angular distance between TLE-predicted and observed sky position of the satellite.
    %\item $\sigma_{r}$ = spatial error of the TLE.
    %\item $\sigma_{t}$ = temporal error of the TLE.
\end{itemize}

\end{appendix}
\end{document}